\documentclass[twocolumn]{aastex631} 
\usepackage{CLASSY_Preamble}

\usepackage{amssymb}	
\usepackage{longtable}
\usepackage{scalerel}

\mathchardef\mhyphen="2D

\newcommand{\oiii}{O\,{\sc iii}}

\newcommand{\siiv}{Si\,{\sc iv}}

\newcommand{\cii}{[C\,{\sc ii}]}

\newcommand{\civ}{C\,{\sc iv}}
\newcommand{\mgii}{Mg\,{\sc ii}}

\newcommand{\angstrom}{\text{ \normalfont\AA}}

\mathchardef\mhyphen="2D

\def\lya{Ly$\alpha$}
\def\ly{$\lambda$}
\def\ha{H$\alpha$}

\def\cii{C\,{\sc ii}}

\def\civ{C\,{\sc iv}}

\def\oi{O\,{\sc i}}
\def\oiii{O\,{\sc iii}}

\def\nai{Na\,{\sc i}}

\def\mgii{Mg\,{\sc ii}}

\def\siiv{Si\,{\sc iv}}
\def\Siii{Si\,{\sc ii}}

\def\siiii{Si\,{\sc iii}}

\def\sii{S\,{\sc ii}}

\def\Q0059{Q0059--2735}

\def\S2S3{S2S3}

\definecolor{blk}{rgb}{0.0,0.0,0.0}
\definecolor{red}{rgb}{0.75,0.0,0.0}
\definecolor{yel}{rgb}{0.65,0.65,0.0}
\definecolor{grn}{rgb}{0.0,0.75,0.0}
\definecolor{blu}{rgb}{0.0,0.0,0.75}
\definecolor{gry}{rgb}{0.75,0.75,0.75}

\def\nh{\ifmmode n_\mathrm{\scriptscriptstyle H} \else $n_\mathrm{\scriptscriptstyle H}$\fi}
\def\ne{\ifmmode n_\mathrm{\scriptstyle e} \else $n_\mathrm{\scriptstyle e}$\fi}
\def\Te{\ifmmode T_\mathrm{\scriptstyle e} \else $T_\mathrm{\scriptstyle e}$\fi}
\def\Qh{\ifmmode Q_\mathrm{\scriptstyle H} \else $Q_\mathrm{\scriptstyle H}$\fi}
\def\Qhesc{\ifmmode Q_\mathrm{\scriptstyle H, esc} \else $Q_\mathrm{\scriptstyle H, esc}$\fi}
\def\Uh{\ifmmode U_\mathrm{\scriptstyle H} \else $U_\mathrm{\scriptstyle H}$\fi}
\def\Nh{\ifmmode N_\mathrm{\scriptstyle H} \else $N_\mathrm{\scriptstyle H}$\fi}
\def\NSi{\ifmmode N_\mathrm{\scriptstyle si} \else $N_\mathrm{\scriptstyle Si}$\fi}
\def\Uhhp{\ifmmode U_\mathrm{\scriptstyle H,HP} \else $U_\mathrm{\scriptstyle H,HP}$\fi}
\def\Nhhp{\ifmmode N_\mathrm{\scriptstyle H,HP} \else $N_\mathrm{\scriptstyle H,HP}$\fi}
\def\Uhvhp{\ifmmode U_\mathrm{\scriptstyle H,VHP} \else $U_\mathrm{\scriptstyle H,VHP}$\fi}
\def\Nhvhp{\ifmmode N_\mathrm{\scriptstyle H,VHP} \else $N_\mathrm{\scriptstyle H,VHP}$\fi}
\def\Nion{\ifmmode N_\mathrm{\scriptstyle ion} \else $N_\mathrm{\scriptstyle ion}$\fi}
\def\nion{\ifmmode n_\mathrm{\scriptstyle ion} \else $n_\mathrm{\scriptstyle ion}$\fi}

\def\Zsun{\ifmmode {\rm Z}_{\odot} \else Z$_{\odot}$\fi}
\def\Msun{\ifmmode {\rm M}_{\odot} \else M$_{\odot}$\fi}
\def\kms{\ifmmode {\rm km~s}^{-1} \else km~s$^{-1}$\fi}
\def\Lya{\ifmmode {\rm Ly}\alpha \else Ly$\alpha$\fi}
\def\Lyb{\ifmmode {\rm Ly}\beta \else Ly$\beta$\fi}
\def\Lyg{\ifmmode {\rm Ly}\gamma \else Ly$\gamma$\fi}
\def\Lyd{\ifmmode {\rm Ly}\delta \else Ly$\delta$\fi}
\def\neaod{\ifmmode n_\mathrm{\scriptscriptstyle AOD} \else $n_\mathrm{\scriptscriptstyle AOD}$\fi}
\def\necrit{\ifmmode n_\mathrm{\scriptstyle cr} \else $n_\mathrm{\scriptstyle cr}$\fi}
\def\ncr{\ifmmode n_\mathrm{\scriptstyle cr} \else $n_\mathrm{\scriptstyle cr}$\fi}
\def\nepi{\ifmmode n_\mathrm{\scriptscriptstyle PI} \else $n_\mathrm{\scriptscriptstyle PI}$\fi}
\def\gtorder{\mathrel{\raise.3ex\hbox{$>$}\mkern-14mu\lower0.6ex\hbox{$\sim$}}}
\def\ltorder{\mathrel{\raise.3ex\hbox{$<$}\mkern-14mu\lower0.6ex\hbox{$\sim$}}}

\def\vro{\ifmmode v_\mathrm{\scriptscriptstyle 1, \scriptstyle r} \else $v_\mathrm{\scriptscriptstyle 1, \scriptstyle r}$\fi}
\def\vrc{\ifmmode v_\mathrm{\scriptscriptstyle 2, \scriptstyle r} \else $v_\mathrm{\scriptscriptstyle 2, \scriptstyle r}$\fi}
\def\vzo{\ifmmode v_\mathrm{\scriptscriptstyle 1, \scriptstyle z} \else $v_\mathrm{\scriptscriptstyle 1, \scriptstyle z}$\fi}
\def\vzc{\ifmmode v_\mathrm{\scriptscriptstyle 2, \scriptstyle z} \else $v_\mathrm{\scriptscriptstyle 2, \scriptstyle z}$\fi}

\newcommand{\Vout}{$V_\text{out}$}

\newcommand{\Rout}{$r_\text{out}$}
\newcommand{\Rphot}{$r_\text{phot}$}
\newcommand{\Rram}{$r_\text{ram}$}
\newcommand{\Rstar}{$r_{*}$}

\newcommand{\Rn}{$r_n$}

\newcommand{\Ncloud}{$N_\text{cl}$}
\newcommand{\Rcloud}{$R_\text{cl}$}
\newcommand{\Mcloud}{$M_\text{cl}$}
\newcommand{\Vturb}{$V_\text{turb}$}
\newcommand{\tcool}{$t_\text{cool}$}
\newcommand{\tmix}{$t_\text{mix}$}
\newcommand{\rcrit}{$R_{crit}$}

\newcommand{\nout}{\ne}

\newcommand{\RC}{R$_\text{Cloud}$}
\newcommand{\MC}{M$_\text{Cloud}$}
\newcommand{\El}{E$_\text{low}$}
\newcommand{\Eu}{E$_\text{up}$}
\newcommand{\ns}{$n_\text{s}$}
\newcommand{\rs}{$s$}

\newcommand{\bsh}{$\beta_{sh}$}

\newcommand{\PI}{Paper~I}
\newcommand{\PII}{Paper~II}
\newcommand{\PIII}{Paper~III}

\defcitealias{Berg22}{\PI}
\defcitealias{James22}{\PII}
\defcitealias{Xu2022}{\PIII}

\begin{document}

\submitjournal{AASJournal ApJ}
\shortauthors{Xu et al.}
\shorttitle{CLASSY~VI: Density, Structure and Size of Galactic Outflows}

\title{CLASSY VI: \footnote{
Based on observations made with the NASA/ESA Hubble Space Telescope,
obtained from the Data Archive at the Space Telescope Science Institute, which
is operated by the Association of Universities for Research in Astronomy, Inc.,
under NASA contract NAS 5-26555.} The Density, Structure and Size of Absorption-Line Outflows in Starburst Galaxies}

\author[0000-0002-9217-7051]{Xinfeng Xu}
\affiliation{Center for Astrophysical Sciences, Department of Physics \& Astronomy, Johns Hopkins University, Baltimore, MD 21218, USA}

\author[0000-0001-6670-6370]{Timothy Heckman}
\affiliation{Center for Astrophysical Sciences, Department of Physics \& Astronomy, Johns Hopkins University, Baltimore, MD 21218, USA}

\author[0000-0002-6586-4446]{Alaina Henry}
\affiliation{Center for Astrophysical Sciences, Department of Physics \& Astronomy, Johns Hopkins University, Baltimore, MD 21218, USA}
\affiliation{Space Telescope Science Institute, 3700 San Martin Drive, Baltimore, MD 21218, USA}

\author[0000-0002-4153-053X]{Danielle A. Berg}
\affiliation{Department of Astronomy, The University of Texas at Austin, 2515 Speedway, Stop C1400, Austin, TX 78712, USA}

\author[0000-0002-0302-2577]{John Chisholm}
\affiliation{Department of Astronomy, The University of Texas at Austin, 2515 Speedway, Stop C1400, Austin, TX 78712, USA}

\author[0000-0003-4372-2006]{Bethan L. James}
\affiliation{AURA for ESA, Space Telescope Science Institute, 3700 San Martin Drive, Baltimore, MD 21218, USA}

\author[0000-0001-9189-7818]{Crystal L. Martin}
\affiliation{Department of Physics, University of California, Santa Barbara, Santa Barbara, CA 93106, USA}

\author[0000-0001-6106-5172]{Daniel P. Stark}
\affiliation{Steward Observatory, The University of Arizona, 933 N Cherry Ave, Tucson, AZ, 85721, USA}


\author[0000-0001-8587-218X]{Matthew Hayes}
\affiliation{Stockholm University, Department of Astronomy and Oskar Klein Centre for Cosmoparticle Physics, AlbaNova University Centre, SE-10691, Stockholm, Sweden}

\author[0000-0002-2644-3518]{Karla Z. Arellano-C\'{o}rdova}
\affiliation{Department of Astronomy, The University of Texas at Austin, 2515 Speedway, Stop C1400, Austin, TX 78712, USA}

\author[0000-0002-9136-8876]{Cody Carr}
\affiliation{Minnesota Institute for Astrophysics, University of Minnesota, 116 Church Street SE, Minneapolis, MN 55455, USA}

\author[0000-0002-9136-8876]{Mason Huberty}
\affiliation{Minnesota Institute for Astrophysics, University of Minnesota, 116 Church Street SE, Minneapolis, MN 55455, USA}

\author[0000-0003-2589-762X]{Matilde Mingozzi}
\affiliation{Space Telescope Science Institute, 3700 San Martin Drive, Baltimore, MD 21218, USA}

\author[0000-0002-9136-8876]{Claudia Scarlata}
\affiliation{Minnesota Institute for Astrophysics, University of Minnesota, 116 Church Street SE, Minneapolis, MN 55455, USA}

\author[0000-0001-6958-7856]{Yuma Sugahara}
\affiliation{National Astronomical Observatory of Japan, 2-21-1 Osawa, Mitaka, Tokyo 181-8588, Japan}
\affiliation{Waseda Research Institute for Science and Engineering, Faculty of Science and Engineering, Waseda University, 3-4-1, Okubo, Shinjuku, Tokyo 169-8555, Japan}

\correspondingauthor{Xinfeng Xu} 
\email{xinfeng@jhu.edu}


\begin{abstract}

Galaxy formation and evolution are regulated by the feedback from galactic winds. Absorption lines provide the most widely available probe of winds. However, since most data only provide information integrated along the line-of-sight, they do not directly constrain the radial structure of the outflows. In this paper, we present a method to directly measure the gas electron density in outflows (\ne), which in turn yields estimates of outflow cloud properties (e.g., density, volume filling-factor, and sizes/masses). We also estimate the distance ($r_n$) from the starburst at which the observed densities are found. We focus on 22 local star-forming galaxies primarily from the COS Legacy Archive Spectroscopic SurveY (CLASSY). In half of them, we detect absorption lines from fine structure excited transitions of \Siii\ (i.e., \Siii*). We determine \ne\ from relative column densities of \Siii\ and \Siii*, given \Siii*\ originates from collisional excitation by free electrons. We find that the derived \ne\ correlates well with the galaxy's star-formation rate per unit area. From photoionization models or assuming the outflow is in pressure equilibrium with the wind fluid, we get $r_n \sim$ 1 to 2$r_*$ or $\sim$ 5$r_*$, respectively, where $r_*$ is the starburst radius. Based on comparisons to theoretical models of multi-phase outflows, nearly all of the outflows have cloud sizes large enough for the clouds to survive their interaction with the hot wind fluid. Most of these measurements are the first-ever for galactic winds detected in absorption lines and, thus, will provide important constraints for future models of galactic winds.

\end{abstract} 



\section{Introduction} 
\label{sec:intro}

Galactic winds are essential to the evolution of galaxies and the intergalactic medium (IGM). In star-forming galaxies (without accreting black holes), these winds are driven by mass, energy, and momentum supplied by star-formation, in the form of radiation, stellar winds, and supernovae \citep[e.g.,][]{Veilleux05}. The latter two result in the creation of a tenuous and energetic wind fluid that flows out and accelerates existing gas clouds, which are observable as warm to cold outflows \citep[e.g.,][]{Xu22a}. Galactic winds and the outflows they drive are able to transport mass/energy/momentum against the gravitational potential of the hosts. Thus, they have been proposed to explain various feedback effects, e.g., regulating the star formation rate (SFR) of the host galaxy \cite[e.g.,][]{Martin05, Rupke05,Cazzoli14, Heckman16}, chemically enriching the circum-galactic medium (CGM) and IGM \cite[e.g.,][]{Heckman00, Dalcanton07, Martin12, Rubin14, Heckman17b,Chisholm18}, and explaining the ``overcooling problem" in cosmological simulations by reducing the baryon fractions in galactic discs \cite[e.g.,][]{Steidel10, Hopkins12}. Outflows can also clear neutral gas away from young starbursts and regulate the escape of Lyman continuum photons, which is responsible for the cosmic reionization \citep[e.g.,][]{Heckman11, Chisholm17, Hogarth20, Carr21, Saldana-Lopez22}.

In the last few decades, galactic winds and outflows have been intensely studied in the literature, especially in star-forming and starburst galaxies, which commonly host powerful outflows \citep[see reviews in][and references therein]{Heckman17a, Rupke18, Veilleux20, Nguyen23}. Outflows are multi-phase \citep[e.g.,][]{Fluetsch21, Marasco22}, but the most abundant data probe the warm ionized phase. This material is believed to be accelerated by the combined momentum of a hot wind fluid created by stellar ejecta and radiation pressure \citep{Heckman17a}. The two major ways to detect the warm ionized gas are from rest-frame UV absorption lines (e.g., \oi, \Siii, \siiv, and \civ), and optical emission lines (e.g., [\oiii] and \ha). Even though both emission and absorption lines can show kinematic features that represent the outflows, they are thought to arise from different environments \citep[e.g.,][]{Chisholm16a}. Emission lines are weighted towards the denser environments (brightness scales with outflow electron density (\nout) squared), while \ne\ have been found to be $\sim$ 100 -- 1000 cm$^{-3}$ \citep[e.g.,][]{Heckman90, Perna20, Marasco22}. On the contrary, absorption lines trace lower density environments (optical depth scales with \nout). Thus, \cite{Wood15} suggests that UV absorption lines can trace larger-scale galactic outflows and are more reliable tracers of warm gas in starburst-driven outflows.

There exist well-developed ways to constrain various important outflow parameters from the absorption lines, including outflow velocity (\Vout), ionization, and column density (\Nh) \citep[e.g.,][]{Martin05, Rupke05, Chisholm15, Scarlata15, Chisholm16a, Heckman16, Carr18, Xu22a}. The strength of the absorption outflows and their potential feedback effects can then be quantified by their mass, momentum, and energy rates, i.e., $\dot{M}_\text{out}\propto\ N_\text{H} r_\text{out} V_\text{out}$, $\dot{p}_\text{out} \propto\ N_\text{H} r_\text{out} V_\text{out}^{2}$, and $\dot{E}_\text{out} \propto \frac{1}{2}N_\text{H} r_\text{out} V_\text{out}^{3}$, respectively, where $r_\text{out}$ is the assumed radius of the outflows. 

The major uncertainty for these outflow rates is from \Rout\ \citep[e.g.,][]{Chisholm16b}. This is because these surveys of galactic outflows in absorption lines only have integrated spectra in a single aperture. It is not possible to measure \Rout\ directly. Most previous studies either assume a fiducial radius \citep[e.g., 1 -- 5 kpc in][]{Martin05, Rupke05, Martin12}, or assume \Rout\ starts at a few times the starburst radius \citep[\Rstar, e.g.,][]{Chevalier85, Heckman15, Chisholm17, Carr21, Xu22a}. Recently, \cite{Wang20} showed that the non-resonant emission lines are much weaker and narrower than the corresponding absorption lines in a sample of starburst galaxies. They suggest that observed absorbing material for outflows could be located at radii significantly larger than \Rstar. 

Moreover, the meaning of \Rout\ is only well-defined for the idealized case in which the outflow is a thin bubble. In the more general case where the outflow is continuous (i.e., \Rout\ is a distribution), the appropriate value of \Rout\ for calculating outflow rates will depend upon the radial variation of density and velocity in the outflow. Without knowing the radial structure of the outflow, outflow rates are uncertain.

In addition to uncertainties in the radial structure of the outflows probed by absorption lines, there is the long-standing theoretical problem about the nature of outflows. How can the absorbing material survive long enough to be accelerated to hundreds of km/s without being shredded by the hydro-dynamical interaction with the wind fluid \citep[e.g.,][]{Nguyen23}? Recent work \citep[e.g.,][]{Gronke20, Fielding22} imply that clouds exposed to an outflowing hot wind can either grow by accreting gas at the cloud's interface with the hot phase (for large clouds), or be destroyed (for small clouds). To date, there are no good empirical constraints on the cloud masses (\Mcloud) or radii ($R_\text{cl}$) in outflows.

In this paper, we aim to shed light on a method to measure \nout, the radius at which these densities apply ($r_n$), \Mcloud, and $R_\text{cl}$ from outflow absorption lines. We focus on 22 local star-forming galaxies selected from the COS Legacy Archive Spectroscopy SurveY (CLASSY) atlas \citep[][]{Berg22, James22} and \cite{Heckman15}. These galaxies have high signal-to-noise ratio (SNR) HST/COS spectra which cover their rest-frame UV bands. In half of the galaxies, we can securely detect absorption lines from the fine structure excited transitions of \Siii, i.e., \Siii*. From it, we determine various important physical parameters of the outflows, including \nout, \Rn, outflow volume filling factor, outflow cloud sizes, and cloud masses. Since the majority of these measurements are the first-ever for galactic winds detected in absorption lines, we discuss their implications and what observational constraints they provide for future models of galactic winds.

The structure of the paper is as follows. In Section \ref{sec:obs}, we introduce the data and observations that are used in this paper. We then describe how to measure the column density from \Siii\ and \Siii*\ in Section \ref{sec:analyses} and how to derive \nout\ from these two quantities. In Section \ref{sec:results}, we present results for the \nout\ and \Rn. We compare them with empirical estimates that are commonly adopted in the literature. We also describe how to derive several other important outflow parameters (the cloud masses, radii, and volume filling factors). Finally, in Section \ref{sec:Discussion}, we discuss and compare our results with other outflow density and radius measurements in the literature. We also contrast our results with current outflow models in Section \ref{sec:Discussion}. We conclude the paper in Section \ref{sec:Conclusion}.

We adopt a cosmology with H$_{0}$ = 69.6 km s$^{-1}$ Mpc$^{-1}$, $\Omega_m$ = 0.286, and $\Omega_{\Lambda}$ = 0.714 \citep{Bennett14}, and we use Ned Wright's Javascript Cosmology Calculator website \citep{Wright06}. In this paper, we adopt the notation $r$ for distances from the starburst, and use $R$ to represent outflow cloud radii.

\section{Observations and Data Reductions}
\label{sec:obs}

In this paper, we select galaxies from the parent sample of the CLASSY dataset \citep{Berg22}, which includes 45 local star-forming galaxies (0.002 $<$ z $<$ 0.182). These galaxies are observed by the G130M+G160M+G185M/G225M gratings on Hubble Space Telescope (HST)/Cosmic Origins Spectrograph (COS) for their rest-frame far-ultraviolet (FUV) spectral regions. To enlarge the dynamic range of the sample at the highest star-formation rates (SFR), we also include five similar galaxies from the Lyman Break Analog (LBA) sample in \cite{Heckman15}. These galaxies have similar quality HST/COS observations as CLASSY ones. We then apply three selection criteria: 1) the SNR per resolution element (0.18\angstrom) in the continuum near 1260 \AA\ in the rest-frame is $\geq 5$; 2) the UV half-light radius of the starburst is $<$ 1.5\arcsec\ (so that the COS spectrum represents the majority of the starburst); 3) an outflow has been detected \citep{Xu22a}. These criteria result in a final sample of 22 galaxies.


All data were reduced locally using the COS data-reduction package CalCOS v.3.3.10\footnote{\url{https://github.com/spacetelescope/calcos/releases}}, including spectra extraction and wavelength calibration. We refer readers to \cite{Berg22} and \cite{James22} for more details about these data reductions and spectral coaddition procedures. We also apply the same reductions to the LBA galaxies from \cite{Heckman15}. Therefore, the whole sample was reduced and processed in a self-consistent way. We have re-sampled the spectra into bins of 0.18\angstrom\ (spectral resolution $\sim$ 6000 -- 10000 from the blue to red end) \citep{Xu22a}. These galaxies' redshift are derived from fitting the optical emission lines discussed in \cite{Mingozzi22}.

\section{Analyses}
\label{sec:analyses}

\subsection{Summary of Previous Outflow Analyses}
\label{sec:outflowSummary}

For our sample, the detailed analyses of outflow properties and their relationship to the host galaxy properties are reported in \cite{Xu22a}. We briefly summarize the key steps as follows.

\begin{enumerate}
    \item Given the reduced data from CalCOS, we start with fitting the stellar continuum of galaxies using stellar models from Starburst99 \citep{Leitherer99, Leitherer10}. We follow the methodology discussed in \cite{Chisholm19}. We then normalize the spectra by the best-fit stellar continuum for each galaxy.
    
    \item For each galaxy, the final reduced HST/COS spectra cover $\sim$ 1200\angstrom\ -- 2000\angstrom\ in the observed frame. In this region, various lines from galactic outflows are detected as absorption troughs, from, e.g., \oi\ \ly 1302, \cii\ \ly 1334, \Siii\ multiplet (\ly1190, 1193, 1260, 1304, and 1526), \siiii\ \ly 1206, and \siiv\ \ly\ly 1393, 1402.
    
    \item To isolate the outflowing gas component from the static ISM, we fit a double-Gaussians model to each absorption trough. The first Gaussian has a fixed velocity center at $v$ = 0 km s$^{-1}$, which represents the static ISM component, and the second Gaussian has a velocity center $<$ 0 km s$^{-1}$, which stands for the blueshifted outflow component. Since the line-spread-functions (LSF) from HST website\footnote{\url{https://www.stsci.edu/hst/instrumentation/cos/performance/spectral-resolution}} is only suitable for point sources, we have constructed non-point source line-spread-functions (LSF) for each galaxy and convolved them with standard Gaussian profiles in the fitting process.
    
    
    \item To robustly measure the ionic column density (\Nion) of outflows, we apply partial coverage (PC) models to \Siii\ multiplet and \siiv\ doublet absorption troughs. From the PC models, we have determined the optical depths, covering fraction (CF), and \Nion\ for \Siii\ and \siiv\ as functions of velocity.
    
    \item We then compare the measured \Nion\ to grids of photoionization models from CLOUDY [version c17.01, \citep{Ferland17}] to determine the total silicon and hydrogen column densities, i.e., N(Si) and \Nh, respectively.
    
    \item We derive the mass/momentum/energy outflow rates given the derived \Nh\ and \Vout, while we assume \Rout\ = \Rstar, which we take to be the radius enclosing 90\% of the starburst FUV emission.

\end{enumerate}

\begin{figure*}
\center
	\includegraphics[angle=0,trim={0.0cm 0.0cm 0.0cm 0.5cm},clip=true,width=0.45\linewidth,keepaspectratio,page=2]{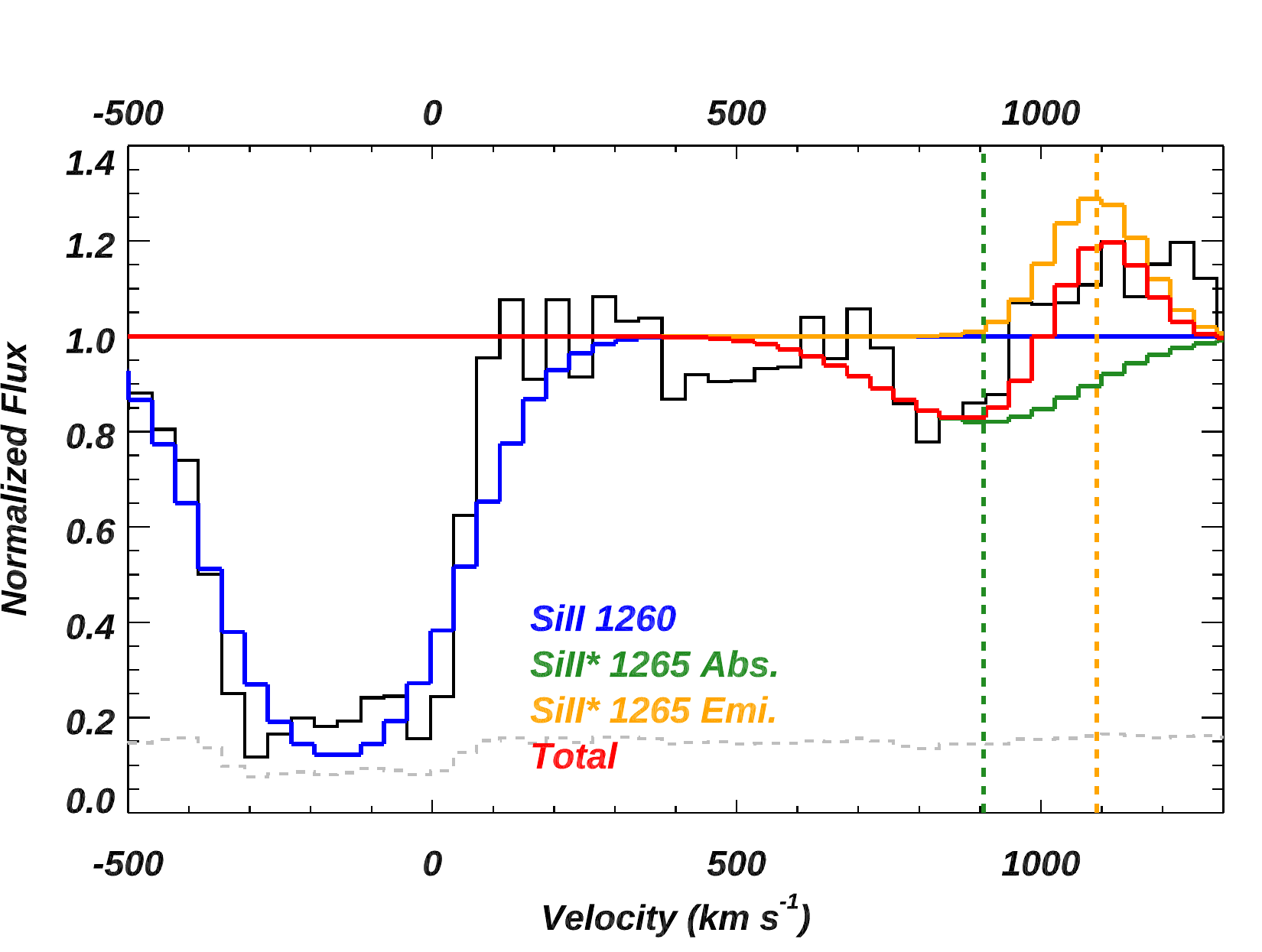}
	\includegraphics[angle=0,trim={0.0cm 0.0cm 0.0cm 0.5cm},clip=true,width=0.45\linewidth,keepaspectratio,page=3]{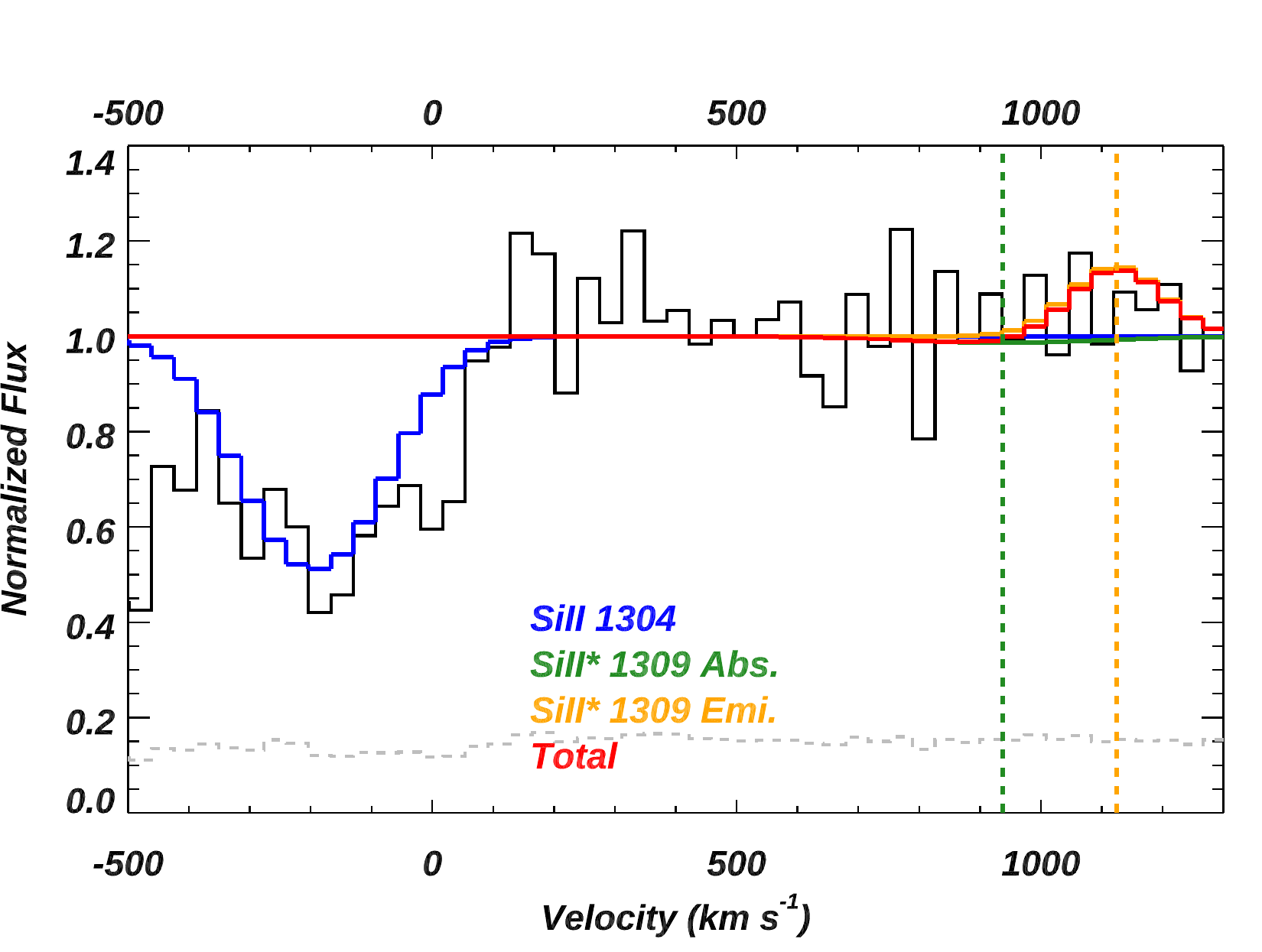}
	
\caption{\normalfont{Example of fitting to the absorption and emission lines for \Siii*\ spectral regions for galaxy J0150+1308 (z = 0.14668). The black and gray histograms are the data and errors, respectively. In each panel, the blueshifted outflow component for the \Siii\ resonance line is shown in blue \citep[adopted from][see Section \ref{sec:outflowSummary}]{Xu22a}. The fitted absorption and emission models for \Siii*\ are shown in green and orange, respectively. The summed model for \Siii*\ is shown in red. The green and orange dashed lines mark the velocity centers for the fitted absorption and emission lines, respectively. See detailed fitting methods in Section \ref{sec:mea}.  } }
\label{fig:SiIIFits}
\end{figure*}

\subsection{Measurements of Column Density from \Siii*}
\label{sec:mea}

Galactic outflows not only show absorption lines from resonance transitions, e.g., \Siii\ \ly 1260, but also from fine structure excited transitions, e.g., \Siii*\ \ly 1265 \citep[e.g.,][]{Jaskot19}. The combination of both can be adopted to derive the electron number density (\ne) of the outflows (see Section \ref{sec:mechan}). In this subsection, we focus on measuring N(\Siii*) for galaxies in our sample.


There are a total of six \Siii*\ lines observable in the rest-frame FUV. We list their important atomic information in Table \ref{tab:atomic}. We find the observed absorption lines from \Siii*\ are commonly weak in our galaxies. This is consistent with the assumed low \ne\ ($\sim$ 10 cm$^{-3}$) for typical starburst galaxies \citep[e.g.,][]{Xu22a}. This low \ne\ has both pros and cons for our analysis. On the one hand, the weaker \Siii*\ troughs are generally optically thin ($\tau\ \ll$ 1), and we can safely measure N(\Siii*) by adopting CF = 1, given the apparent optical depth (AOD) assumption \citep{Savage91}. On the other hand, the shallow \Siii*\ troughs are sometimes difficult to measure, even in our high SNR HST/COS spectra. Another complexity is that the emission lines from \Siii*\ (i.e., so-called fluorescent lines) can contaminate the blue-shifted absorption troughs of \Siii*, especially when the outflow velocity (\Vout) is small. The steps for our fitting process of \Siii*\ lines and measurements of N(\Siii*) are as follows:


\begin{enumerate}
    \item  We fit the fluorescent emission and fine-structure absorption lines from \Siii* \ly 1197, 1265, 1309, and 1533 for each galaxy simultaneously. We exclude \Siii*\ \ly 1194 because it is commonly blended with the absorption trough from \Siii\ \ly 1193. For each galaxy, we also exclude \Siii*\ lines that fall into a chip gap or are contaminated by Galactic lines (e.g., \Siii\ \ly 1197 can be affected by Galactic \lya).

    \item For each \Siii*\ absorption line, we assume it has a Gaussian optical depth profile:
    
\begin{equation}
\label{Eq:GaussianFit1}
    \begin{aligned}  
        I_{k}(v) &= e^{-\tau_{k}(v)}\\
        \tau_{k}(v) &= \frac{b_{k}}{\sigma\sqrt{2\pi}}\times exp(\frac{(v-v_{c})^{2}}{2\sigma^{2}})
    \end{aligned}  
\end{equation}    
    where $k$ stands for the $k$th \Siii*\ absorption line, $I_{k}(v)$ is the normalized intensity, $\tau_{k}(v)$ is the optical depth at each velocity of the absorption trough, $\mathit{v}$ is the velocity. Given the AOD assumption, the optical depths of different \Siii*\ absorption lines (scaled by coefficient $b_{k}$) are linked by their oscillator strength ($f$) ratios (see Table \ref{tab:atomic}). The velocity center (v$_{c}$) and dispersion ($\sigma$) of the \Siii*\ absorption lines are fixed among all \Siii*\ lines. These fixed values are chosen to be the same as the median values from all \Siii\ resonance absorption lines (Section \ref{sec:outflowSummary}). This assumes that the same outflow clouds have produced \Siii\ and \Siii*\ absorption lines, which is true since both lines have close energy levels (Section \ref{sec:mechan}).

    \item For each \Siii*\ emission line, we model it using only one Gaussian profile in velocity space. This is because \Siii*\ in our sample show weak and narrow fluorescent emission-lines, and are inconsistent with arising from the outflowing gas seen in absorption (i.e., broad lines). This implies that most of the emission from the outflow arises on scales larger than the projected COS aperture \citep{Wang20}. For all \Siii*\ emission lines, we fix v$_{c}$ at the systematic velocity, and $\sigma$ is set to be in the range between 0 and the median FWHM of the static ISM component of \Siii\ resonance lines (Section \ref{sec:outflowSummary}). Their amplitudes are free parameters.

    \item Then, we conduct $\chi^2$ minimization to fit all 2$\times$N profiles simultaneously to the spectral regions of \Siii*. Here the 2 stands for the emission and absorption line for each \Siii*, and N equals the number of \Siii*\ lines that are clean and used in the fit. We adopt the fitting routine \textit{mpfit} \citep{Markwardt09}.
    
    \item Finally, assuming AOD, N(\Siii*) can be derived from the best-fitted $\tau_{k}(v)$ as follows \citep{Savage91}:

\begin{equation}
\label{Eq:GaussianFit2}
	\begin{aligned}
	N_{ion}(v)  &=\frac{3.8 \times\ 10^{14}}{f_{k}\cdot\lambda_{k}}\cdot \tau_{k}(v)\\
	N_{ion}     &= \int {N_{ion}(v)} dv  
	\end{aligned}
\end{equation}
where $\lambda_{k}$ is the wavelength for the $k$th \Siii*\ line that has $\tau_{k}(v)$. Note that under the AOD assumption, choices of different \Siii*\ lines in Equation (\ref{Eq:GaussianFit2}) lead to the exact same \Nion.

\end{enumerate}

\begin{table}
	\centering
	\caption{Atomic Data for the Resonance and Excited Transitions of \Siii\ $^{(a)}$}
	\label{tab:atomic}
	\begin{tabular}{llccl} 
		\hline
		\hline
		Ions & Vac. Wave. & f$_{lk}$ & A$_{kl}$   &\El\ -- \Eu\ \\
		 (1) & (2) & (3) & (4) & (5) \\
		\hline
		\hline
	
		\Siii\  & 1190.42   & 2.77 $\times$ 10$^{-1}$& 6.53 $\times$ 10$^{8}$   & 0.0 - 10.41\\
		\Siii\  & 1193.29   & 5.75 $\times$ 10$^{-1}$& 2.69 $\times$ 10$^{9}$   & 0.0 - 10.39\\
		\Siii\  & 1260.42   & 1.22  & 2.57 $\times$ 10$^{9}$   & 0.0 - 9.84\\
		\Siii\  & 1304.37   & 9.28 $\times$ 10$^{-2}$& 3.64 $\times$ 10$^{8}$   & 0.0 - 9.50\\
		\Siii\  & 1526.71   & 1.33 $\times$ 10$^{-1}$& 3.81 $\times$ 10$^{8}$   & 0.0 - 8.12\\
		\hline
		\Siii*\  & 1194.50   & 7.37 $\times$ 10$^{-1}$& 3.45 $\times$ 10$^{9}$   & 0.036 - 10.41\\
		\Siii*\  & 1197.39   & 1.50 $\times$ 10$^{-1}$& 1.40 $\times$ 10$^{9}$   & 0.036 - 10.39\\
		\Siii*\  & 1264.73$^{(b)}$   & 1.09  & 3.04 $\times$ 10$^{9}$   & 0.036 - 9.84\\
		\Siii*\  & 1265.02$^{(b)}$   & 1.13$\times$ 10$^{-1}$  & 4.73 $\times$ 10$^{8}$   & 0.036 - 9.84\\
		\Siii*\  & 1309.28   & 8.00 $\times$ 10$^{-2}$& 6.23 $\times$ 10$^{8}$   & 0.036 - 9.50\\
		\Siii*\  & 1533.45   & 1.33 $\times$ 10$^{-1}$& 7.52 $\times$ 10$^{8}$   & 0.036 - 8.12\\		

		\hline
		\hline
	\multicolumn{5}{l}{%
  	\begin{minipage}{8cm}%
	Note. --\\
	    \textbf{(a).}\ \ Data are obtained from National Institute of Standards and Technology (NIST) atomic database \citep{Kramida18}. \\
    	\textbf{(2).}\ \ Vacuum wavelengths in units of \angstrom.\\
        \textbf{(3).}\ \ Oscillator strengths.\\
        \textbf{(4).}\ \ Einstein A coefficients in units of s$^{-1}$.\\
        \textbf{(5).}\ \ Energies from lower to upper levels in units of eV.\\
        \textbf{(b).}\ \ \Siii*\ has two close transitions at $\sim$ 1265\angstrom, i.e., \Siii*\ \ly1264.73 and \ly1265.02. Both are from the same lower energy level at 0.036 eV, but have slightly different upper energy levels due to fine structure splitting ($\delta E$ $\sim$ 4$\times$10$^{-4}$ eV). See discussion in Section \ref{sec:mechan}.\\
  	\end{minipage}%
	}\\
	\end{tabular}
	\\ [0mm]
	
\end{table}

There are two close transitions of \Siii*\ at $\sim$ 1265\angstrom, i.e., \Siii*\ \ly1264.73 and \ly1265.02. Both are from the same lower energy level at 0.036 eV (= 287.24 cm$^{-1}$), but have slightly different upper energy levels due to fine structure splitting ($\delta E$ $\sim$ 4$\times$10$^{-4}$ eV). Since the velocity offset between these two lines is only 69 km s$^{-1}$, we can barely resolve their absorption lines in the spectra. Thus, we adopt the combined $f$ value in the calculations of N(\Siii*) \citep{Borguet12}. Since \Siii*\ \ly1265.02 has $\sim$ 10 times smaller $f$ value than that of \Siii*\ \ly1264.73 (Table \ref{tab:atomic}), the absorption trough is 
always dominated by the latter.

An example of the fitted \Siii*\ spectral regions is shown in Figure \ref{fig:SiIIFits}. The outflow components for the \Siii\ resonance lines are shown in blue (Section \ref{sec:outflowSummary}), while the fitted absorption and emission models for \Siii*\ are shown in green and orange, respectively. The overall model for \Siii*\ by summing both the absorption and emission is shown in red. There is a clear absorption trough from \Siii*\ \ly1265, and it is well-fitted, while there is no trough seen in \Siii*\ \ly1309. This is as expected since f$_{1265}$/f$_{1309}$ = 15, which leads to $\tau_{1265}$/$\tau_{1309}$ = 15 under AOD models. Overall, we have measured N(\Siii*) securely in 11 out of 22 galaxies in our sample. These galaxies that have N(\Siii*) measured yield a mean N(\Siii*)/N(\Siii) $\sim$ 0.01. In Table \ref{tab:mea}, we report the measured N(\Siii) and N(\Siii*) in column 6 and 7, respectively. 



\subsection{Mechanisms for Generating \Siii*: Collisions v.s. Radiative Pumping}
\label{sec:mechan}

As shown in Table \ref{tab:atomic}, the observed fine-structure transitions of \Siii*\ in FUV have lower energy levels as \El\ = 0.036 eV (= 287.24 cm$^{-1}$), which is the first excited energy level of \Siii\ (hereafter, \Siii*\ specifically stands for this level). Two mechanisms can populate \Siii*: 1) Collisional excitation of the ground state of \Siii\  by free electrons \citep[e.g.,][]{Silva02, Osterbrock06, Borguet12}. In this case, a higher \ne\ would yield a higher n(\Siii*)/n(\Siii) ratio, where n(\Siii*) and n(\Siii) stand for the level population of the first excited and ground state of \Siii. 2) Indirect UV pumping, i.e., the \Siii\ ground state is excited by absorption of a UV photon to an upper energy level, followed by a spontaneous decay to the excited level at \Siii*\ 287.24 cm$^{-1}$. In this case, a stronger radiation field leads to higher n(\Siii*)/n(\Siii) \citep[see, e.g.,][]{Prochaska06}.



To check if indirect UV pumping can be the dominant mechanism, we estimate the radiation intensity G (in units of ergs cm$^{-2}$ s$^{-1}$) suffered by outflows for galaxies in our sample. We first measure each galaxy's continuum flux ($\lambda$F$_{\lambda}$) around \Siii\ \ly 1260. Then we convert it to luminosity as:  $\lambda$L$_{\lambda}$ = 4$\pi$D$_{L}^{2} \times\lambda$F$_{\lambda}$, where D$_{L}$ is the luminosity distance of the galaxy. We conservatively assume the location of observed outflowing gas is at or beyond the starburst radius (i.e., \Rout\ $>$ \Rstar), which we will show in Section \ref{sec:R} to be a fair assumption. Finally, we derive G for each galaxy as G = $\lambda$L$_{\lambda}$/(4$\pi r_\text{out}^{2}$). We find the majority of galaxies in our sample have G/G$_{0}$ $<$ 10$^{3}$ (with two exceptions), while the mean G/G$_{0}$ is only $\sim$ 250. Here, G$_{0}$ represents the interstellar FUV intensity of our Milky Way \citep{Habing68}, which is $\sim$ 1.6 $\times$ 10$^{-3}$ ergs cm$^{-2}$ s$^{-1}$. 


As shown in \cite{Prochaska06}, n(\Siii*)/n(\Siii) $<$ 10$^{-4}$ when G/G$_{0}$ $<$ 10$^{3}$. Given our observed mean N(\Siii*)/N(\Siii) = 0.01\footnote{In the LOS, the observed N(\Siii*)/N(\Siii) = n(\Siii*)/n(\Siii).}, we conclude that indirect UV pumping commonly contribute $<$ 10$^{-4}$/0.01 = 1\% of the observed population of \Siii*. This is different from the fine structure absorption lines detected in $\gamma$-ray bursts in \cite{Prochaska06}, where indirect UV pumping dominates because the radiation field is much stronger. 





\begin{figure}
\center
	\includegraphics[page = 1, angle=0,trim={0.1cm 0.1cm 0.1cm 0.3cm},clip=true,width=1.0\linewidth,keepaspectratio]{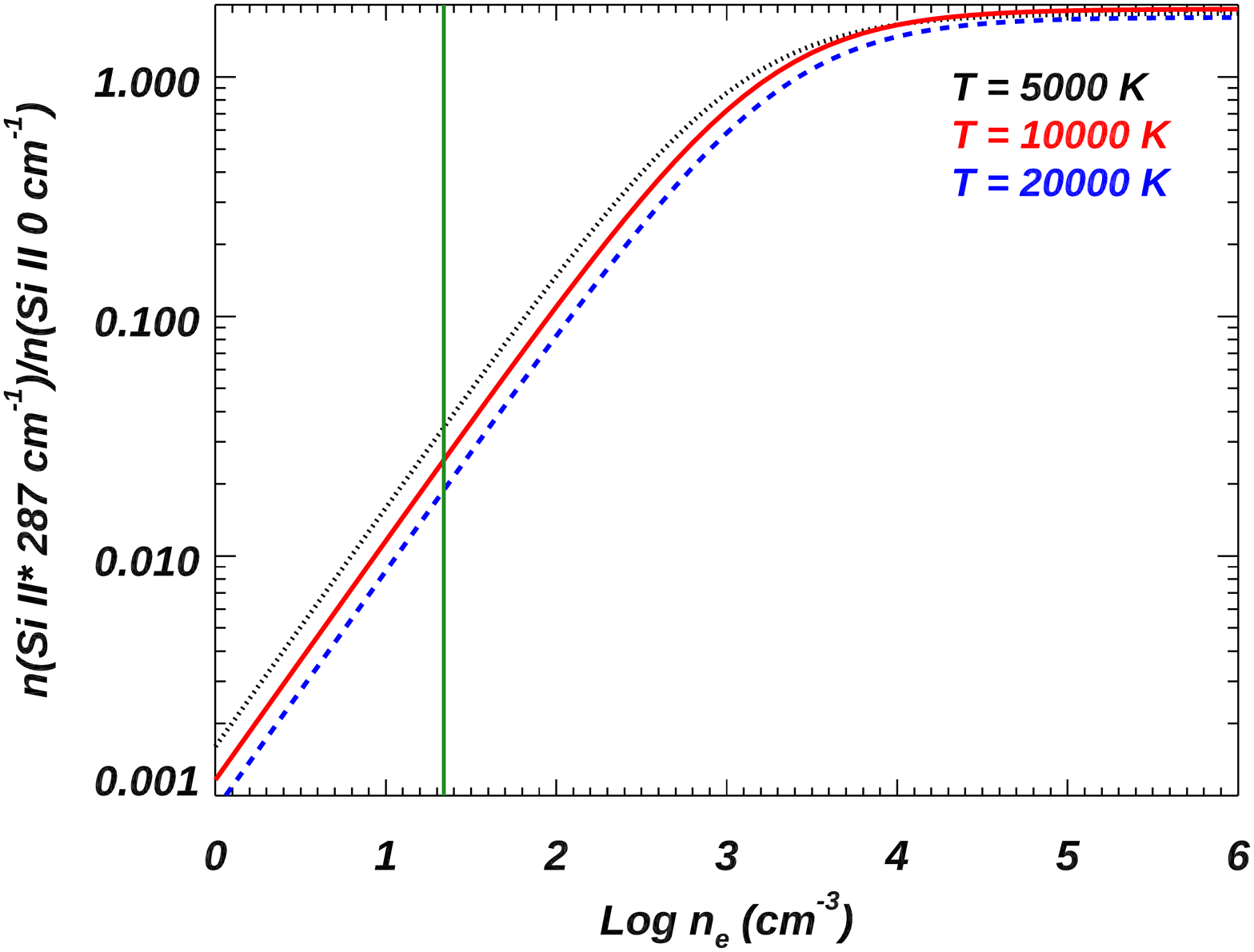}

\caption{\normalfont{Population ratio of \Siii's fine structure level (\El\ = 287 cm$^{-3}$) to the ground state (\El\ = 0 cm$^{-3}$) versus the electron number density (\ne) \citep[e.g.,][]{Osterbrock06}. The modelled curves are calculated using the CHIANTI database \citep[v8.0.7,][]{DelZanna15}, assuming collisional excitation under three different temperatures. The green vertical line represents the median value of \ne\ measured from galaxies in our sample. See Section \ref{sec:mechan} for more discussion. } }
\label{fig:ne}
\end{figure}

Thus, collisional excitation is the dominant mechanism for populating n(\Siii*) in our galaxies. We show the relationship between level population ratio and \ne\ in Figure \ref{fig:ne} for \Siii. The modelled curves are calculated using the CHIANTI database \citep[v8.0.7,][]{DelZanna15}, assuming collisional excitation under three different temperatures. The relation is only weakly dependent on temperature. The critical density (\necrit) for \Siii*\ is defined at the position where $n$(\Siii*) = $n$(\Siii). For T = 10,000 K, we get \necrit\ $\sim$ 2000 cm$^{-3}$. From Figure \ref{fig:ne}, we can derive \ne\ from the observed column density ratio of N(\Siii*)/N(\Siii) \citep[e.g.,][]{Borguet12, Xu19}. The errors of \ne\ are propagated from the errors of N(\Siii*)/N(\Siii).

Given galaxies in our sample have N(\Siii*)/N(\Siii) in the range between $\sim$ 0.001 and $\sim$ 0.1, we get \ne\ from a few to $\sim$ 100 cm$^{-3}$. The derived \ne\ for each galaxy is listed in Table \ref{tab:mea}. For galaxies that show no absorption on N(\Siii*), we present upper limits on \ne\ based on their upper limits of N(\Siii*) (Section \ref{sec:mea}).

\begin{table*}
	\centering
	\caption{Summary of the Notations and Measured Quantities}
	\label{tab:stat}
	\begin{tabular}{llllll} 
		\hline
		\hline
		Notation & Definition & Reference & Mean & Median & STDDEV$^{(a)}$  \\
		 (1) & (2) & (3) & (4) & (5) & (6)\\
		\hline
		\hline

		\Rout\   & Actual outflow radius distribution   & Section \ref{sec:intro}     & \dots$^{(b)}$  & \dots  & \dots\\
    	\Rstar\   & Starburst radius of the galaxy           & Section \ref{sec:intro}     & \dots & \dots  & \dots\\
		$r_{n}$\   & Outflow radius at the derived \ne\ from \Siii*           & Section \ref{sec:Inter}     & \dots$^{(c)}$ & \dots  & \dots\\

             \hline
             
		\ne\      & Outflow electron number density           & Section \ref{sec:ne_Corr} & 22.3 cm$^{-3}$ & 22.8 cm$^{-3}$   & 18.7 cm$^{-3}$\\
		\Rphot\   & Outflow radius assuming photoionization           & Section \ref{sec:R_U}     & 1.6 kpc & 1.2 kpc   & 1.4 kpc\\
		\Rram\    & Outflow radius assuming pressure equilibrium       & Section \ref{sec:R_P}     & 4.2 kpc & 4.1 kpc   & 2.0 kpc\\
		FF\       & Outflow volume filling factor             & Section \ref {sec:other}  & 0.5\%     & 0.4\% & 0.4\% \\
  		\Rcloud\  & Outflow cloud size   & Section \ref {sec:other}      & 13 pc & 5 pc   & 15 pc\\
    	\Mcloud\  & Outflow cloud mass  & Section \ref {sec:other}      & 1.1 $\times$ 10$^{4}$ \Msun & 202 \Msun   & 1.0 $\times$ 10$^{4}$ \Msun\\

		\hline
		\hline
	\multicolumn{6}{l}{%
  	\begin{minipage}{15cm}%
	Note. -- \\
           \textbf{(*).}\ \ The first part of this table are the important notations adopted throughout the paper. The second part (beginning with \ne) shows the measured quantities (galaxies with lower and upper limits are excluded). We show the mean, median values, and standard deviations for these quantities. \\
      	\textbf{(a).}\ \ Standard deviation. \\
      	\textbf{(b).}\ \ \Rout\ is a range or distribution, which can only be measured from spatially resolved detections of outflows. \Rout\ = $r_{n}$ only when the outflow is a thin bubble (see discussions in Sections \ref{sec:intro} and \ref{sec:R}).   \\
      	\textbf{(c).}\ \ Based on different assumptions, we can measure $r_{n}$ as \Rphot\ or \Rram\ specifically (illustrated in Section \ref{sec:R}).  \\
       
  	\end{minipage}%
	}\\
	\end{tabular}
	\\ [0mm]
	
\end{table*}

\section{Results}
\label{sec:results}

We summarize the main notations and measured quantities in this paper at Table \ref{tab:stat}. We illustrate their details in the following subsections.

\begin{figure}
\center
    \includegraphics[page = 1, angle=0,trim={0.1cm 0.1cm 0.1cm 0.3cm},clip=true,width=1.00\linewidth,keepaspectratio]{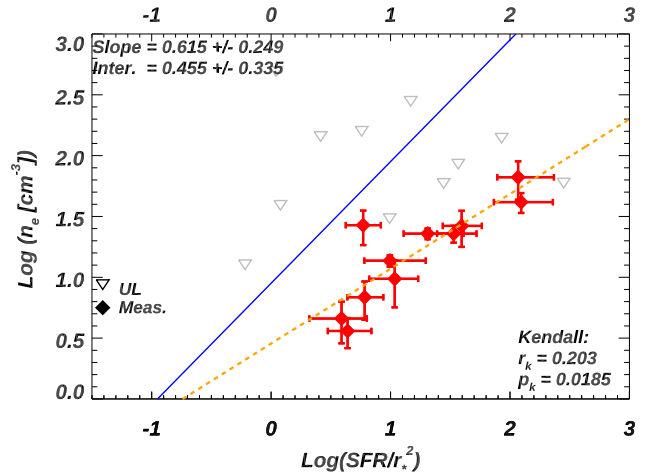}
\caption{\normalfont{Strong orrelations between outflow electron number density (\ne) and SFR surface density. Galaxies that have \ne\ measurement (Mea.) or upper limits (UL) are shown as the red-filled or gray-open symbols, respectively. Kendall's $\tau$ correlation coefficients are shown at the bottom-right corner of each panel, where we have considered the upper limits following \cite{Akritas96}. The best linear-fit to all measurements is shown as the orange dashed line, and the fitted slope and intercept are shown in the top-left corner. The blue line represents the model from \cite{Chevalier85} assuming the outflow gas is in pressure balance with the wind fluid at the radius of the starburst [Equation (\ref{eq:PressureEqui2}) and Section \ref{sec:R_P}]}. }
\label{fig:ne_Corr}
\end{figure}

\subsection{Outflow Density Distribution and Correlations}
\label{sec:ne_Corr}
In Table \ref{tab:stat}, we show the statistics of the derived \ne\ for 11 galaxies which have secure measurements of their N(\Siii*). We find outflows in the galaxies have the mean and median value \ne\ $\sim$ 23 cm$^{-3}$. These values are consistent with what has been estimated before from absorption-line data for starburst galaxies \citep[e.g., \ne = 19 -- 34 cm$^{-3}$ in][]{Chisholm18}. In Figure \ref{fig:ne_Corr}, we show a strong positive correlation between \ne\ with the SFR surface density. We will discuss the implications of this below.

In all Figures in this Section, galaxies in our sample with \ne\ measurement or upper limits are shown as red-filled or gray-open symbols, respectively. Kendall's $\tau$ correlation coefficient ($r_{k}$) and the probability of the null hypothesis ($p_{k}$) are shown at the bottom-right corner. We have taken account of these upper limits in the Kendall $\tau$ test following \cite{Akritas96}. 

%


\subsection{Interpretations of Outflow Density in Models}
\label{sec:Inter}
Since our HST/COS spectra are integrated over the whole line-of-sight (LOS), the derived \ne\ values for outflows also represent mean values over the velocity profile. To better interpret the measured \ne\ discussed above, we consider  two common outflow models \citep[e.g.,][]{Xu22a} as follows.

The simplest case for outflow is an expanding thin shell model given a mean electron number density (i.e., \ne\ = \ns) and shell thickness (\rs). In this case, we have: 

\begin{equation}\label{eq:NShell}
    \begin{aligned}
        N(\text{Si II*}) &= n(\text{Si II*}) \times s\\
        N(\text{Si II})  &= n(\text{Si II}) \times s
     \end{aligned}
\end{equation}
where \Nion\ and \nion\ represent the column and number density for a certain ion, respectively. For galactic outflows, \ne\ varies between $\sim$ 10 cm$^{-3}$ and $\sim$ 2000 cm$^{-3}$ \citep[e.g.,][]{Chevalier85, Yoshida19}. In this range, the curve in Figure \ref{fig:ne} is approximately linear, so we have:

\begin{equation}\label{eq:neapp}
    \begin{aligned}
        \frac{n(\text{Si II*})}{n(\text{Si II})} \approx \frac{n_\text{e}}{n_\text{cr}}
     \end{aligned}
\end{equation}

Combining Equation (\ref{eq:NShell}) and (\ref{eq:neapp}), we get N(\Siii*)/N(\Siii) $\approx$ \ne/\necrit\ = \ns/\necrit. Therefore, for a thin shell outflow model, the derived \ne\ from N(\Siii*)/N(\Siii) discussed in Section \ref{sec:mechan} is just the mean density in the shell.

In the second case, we consider a mass-conserving galactic wind with constant velocity \citep[e.g.,][]{Carr21}. In this case, the outflow has a density profile $n(r) = n_0 (r/r_{0})^{-2}$, where $r_{0}$ is the radius at which the outflow begins, and $n_0$ is the density at this radius. In this case, we have:

\begin{equation}\label{eq:Nflow1}
    \begin{aligned}
        N(\text{Si II})  &= \int_{r_{0}}^{\infty} n(\text{Si II}) dr = \text{C}_{0} \times n_0 r_{0}
     \end{aligned}
\end{equation}
where the integration is from $r_{0}$ to infinity (note n($\infty$) = 0) and C$_{0}$ = $n(\text{Si II})_{0}/n_{0}$ is the conversion factor from gas number density to \Siii\ number density at $r_{0}$. C$_{0}$ depends on gas metallicity and ionization. Similarly, for \Siii*, we get:

\begin{equation}\label{eq:Nflow2}
    \begin{aligned}
        N(\text{Si II*})  &= \int_{r_{0}}^{\infty} n(\text{Si II*}) dr  \\
                          &\approx \int_{r_{0}}^{\infty} \frac{n(r)^{2}}{n_\text{cr}} \times \text{C(r)} dr
                          &= \frac{C_{0} n^{2}_{0} r_{0}}{3 n_\text{cr}}
     \end{aligned}
\end{equation}
where in the second row we have adopted Equation (\ref{eq:neapp}) to replace $n$(\Siii*). Thus, this mass-conserving outflow model yields N(\Siii*)/N(\Siii) = $n_{0}$/(3\necrit), i.e., the derived \ne\ from Section \ref{sec:mechan} is a third of $n_{0}$. Equivalently, our derived \ne\ corresponds to the gas density at $r_n = \sqrt{3} r_{0}$. Hereafter, we define $r_n$ as the radius of the outflows at which the mean \ne\ derived from fine-structure absorption lines above would occur.


Similarly, if we take a general form of $n(r) = n_0 (r/r_{0})^{-\gamma}$, we get:

\begin{equation}\label{eq:Nflow2}
    \begin{aligned}
        &N(\text{Si II})  = \frac{\text{C}_{0} \times n_0 r_{0}}{(\gamma - 1)}\\
        &N(\text{Si II*}) = \frac{C_{0} n^{2}_{0} r_{0}}{ (2\gamma - 1) n_\text{cr}}\\
        &N(\text{Si II*})/N(\text{Si II}) = \frac{\gamma - 1}{2\gamma -1}\frac{n_0}{n_\text{cr}}
     \end{aligned}
\end{equation}

\begin{figure*}
\center
    \includegraphics[page = 1, angle=0,trim={0.0cm 0.1cm 0.1cm 0.3cm},clip=true,width=0.45\linewidth,keepaspectratio]{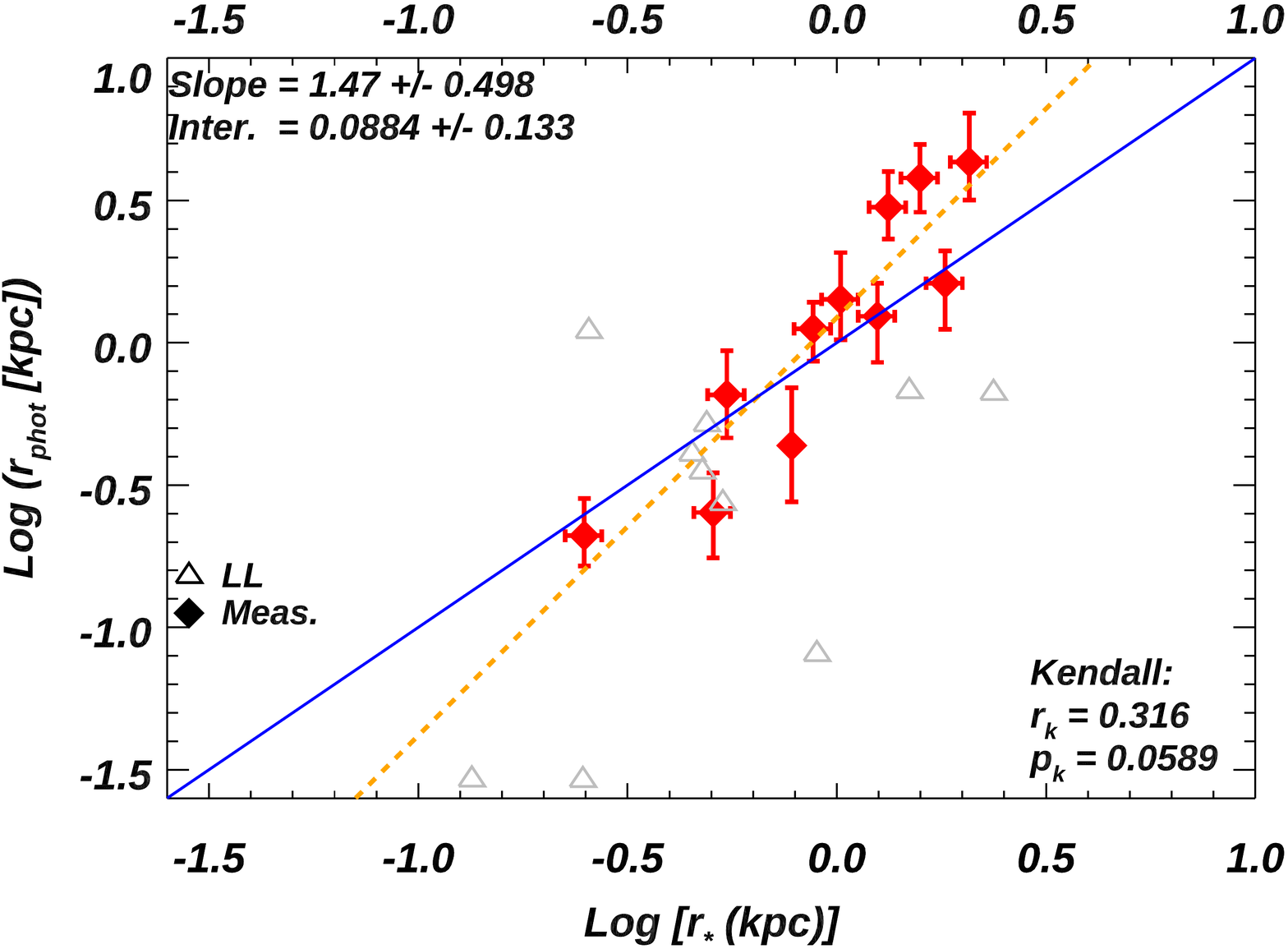}
    \includegraphics[page = 1, angle=0,trim={0.0cm 0.1cm 0.1cm 0.3cm},clip=true,width=0.45\linewidth,keepaspectratio]{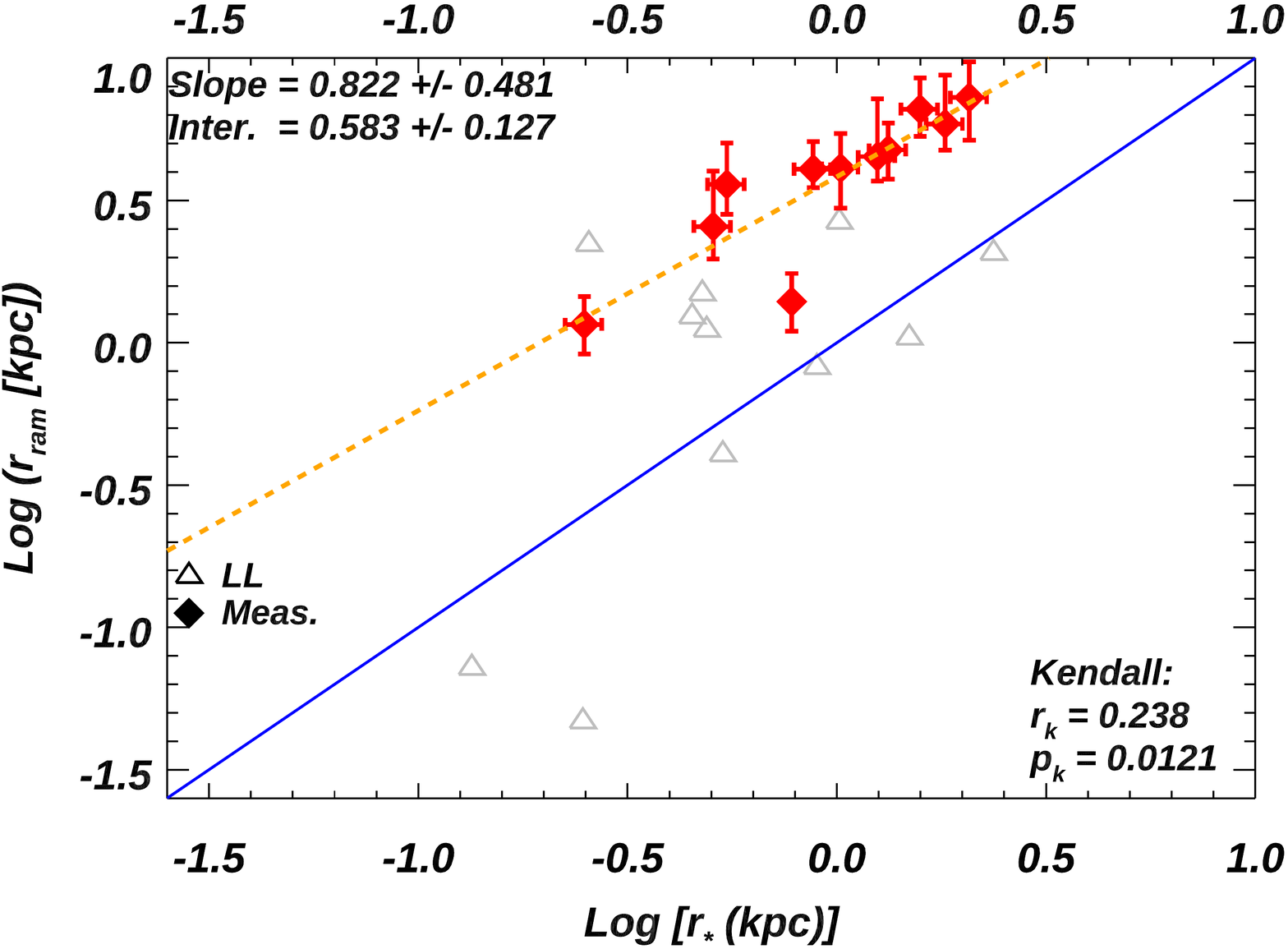}
\caption{\normalfont{Comparisons of the derived outflow distances with the starburst radius ($r_*$)}. The labels and symbols are the same as Figure \ref{fig:ne_Corr}. Galaxies that have $r$ as a measurement or a lower limit are shown as the red-filled or gray-open symbols, respectively. \textbf{Left:} Derived
$r_{phot}$ from photoionization outflow models (Section \ref{sec:R_U}). \textbf{Right:} Derived $r_{ram}$ by assuming pressure equilibrium (Section \ref{sec:R_P}). We show the 1:1 correlation as the blue lines. In both cases, the derived $r$ correlate strongly with $r_*$.}
\label{fig:RComp2}
\end{figure*}

Thus, in the general form, the derived \ne\ from integrated spectra corresponds to $\frac{\gamma - 1}{2\gamma -1}n_{0}$. Equivalently, the derived \ne\ is the gas density at a radius of $(\frac{2\gamma-1}{\gamma-1})^{\frac{1}{\gamma}} \times$ $r_{0}$. Given the evidence for relatively shallow radial density profiles found in outflows \citep[e.g.,][]{Wang20, Burchett21}, we consider the additional cases $\gamma = $ 1.5 and 1.2, and get $r_n = 2.52~r_{0}$ and 5.06~$r_{0}$, respectively.\footnote{Note that these expressions diverge for $\gamma \leq 1$, so we do not consider these shallower profiles.} We will compare these sizes with those estimated from our measured values of \ne\ below. 

We note that if the outflows are more complex than a general form of $n(r) = n_0 (r/r_{0})^{-\gamma}$, or if there is a range in density at a given radius, the exact interpretation of our measured \ne\ and \Rout\ will be different and dependent on the actual form of $n(r)$. We do not dive in this direction, which is beyond the scope of this paper.

\subsection{Derivations of Outflow Distances from \ne}
\label{sec:R}

For galactic outflows, their radial extent (also referred as the ``outflow distance'') can not be determined given only LOS integrated spectra \citep[see, e.g.,][]{Wang20}. In fact, for a continuous outflow, there is no unique way to define distances (e.g., one could define minimum or maximum values, or a half-mass radius, etc.). Moreover, while we have measured a single value for \ne\, for a continuous outflow, this value will only apply at some specific radius in the outflow (i.e., $r_n$). Here we will estimate $r_n$ based on two methods as follows.

\subsubsection{Outflow Distances Assuming Photoionization}
\label{sec:R_U}
In star-forming galaxies, the ultraviolet outflow absorption lines (e.g., from \oi, \Siii, \siiii, \siiv) have been found to be well-described by photoionization models instead of shock-heating models \citep{Chisholm16a}. In this case, we have:

\begin{equation}
\label{eq:Uh}
    \Uh=\frac{\Qh}{4\pi r_\text{photo}^2 \nh c} 	\longrightarrow  r_\text{phot}=\sqrt{\frac{\Qh}{4\pi \Uh \nh c}}
\end{equation}
where \Uh\ is the ionization parameter, \Qh\ is the source emission rate of ionizing hydrogen photons, $c$ is the speed of light, and \nh\ is the hydrogen number density of the outflow. On the right side of the Equation (\ref{eq:Uh}), we show the solved formula for outflow distance $r$ assuming photoionization (hereafter, \Rphot).

For \Qh, we adopt the values from spectral energy distribution (SED) fitting with UV and optical photometry described in \cite{Berg22}. We note that the resulting \Qh\ value is the intrinsic value, but only a portion of the ionizing photons can reach the observed outflows due to attenuation by neutral hydrogen and dust. Thus, we estimate the escaped ionizing photon rate (\Qhesc) as \citep[e.g.,][]{Xu22b}:

\begin{equation}
\label{eq:Qh}
    \Qhesc = Q_\text{H,tot} \times (1 - CF) \times 10^{-0.4E(B-V)k(912)}
\end{equation}
where, for each galaxy, $CF$ represents the covering fraction of the static ISM component derived from the absorption line profiles in \cite{Xu22a}, $E(B-V)$ is the internal dust extinction \citep[derived in][]{Berg22}, and $k(912)$ = 12.87 is the extinction curve at the Lyman limit by assuming the extinction law from \cite{Reddy16b}. The second term on the right of Equation (\ref{eq:Qh}) represents the attenuation by neutral hydrogen, where a fraction of $CF$ around the galaxy is covered by ISM and is generally optically thick to \Qh. The third term stands for the attenuation by dust. We note that this assumes all the extinction arises inside the starburst and that the outflow is at least as large as the starburst. 

For \Uh, we adopt the values determined from outflow absorption lines of \Siii\ and \siiv\ as described in \cite{Xu22a}. For \nh, we approximate it as $\sim$ \ne/1.2, which is applicable for ionized gas, assuming $\sim$ 90\% hydrogen and $\sim$ 9\% helium and some metals. Overall, we can solve \Rphot\ from Equation (\ref{eq:Uh}). The derived results are shown in Table \ref{tab:mea}, which are in the range of 0.2 -- 5 kpc. In the left panel of Figure \ref{fig:RComp2}, we compare the \Rphot\ values with \Rstar, which is the commonly assumed outflow radius in the literature \citep[e.g.,][]{Heckman15, Xu22a}. We see a strong correlation with \Rphot $\sim 1$ to 2\Rstar\ as we move from the smallest to largest galaxies.


\subsubsection{Outflow Distances Assuming Pressure Equilibrium}
\label{sec:R_P}

In this section, we compare our data to the simple analytic model for a starburst-driven wind by \cite{Chevalier85} (CC85). To review, CC85 model assumes that massive stars return mass and kinetic energy to the starburst through supernova explosions and stellar winds. These ejecta are thermalized through shocks to form a very hot region of gas inside the starburst. This gas expands through a sonic radius (the starburst radius \Rstar) and becomes a high-velocity supersonic wind that can accelerate clouds in its path, producing the blue-shifted absorption lines we see. This latter gas is much denser and cooler than the wind fluid.

The CC85 model requires a density of the wind fluid at its sonic point that depends on SFR/$r_*^2$. We see this dependence for the absorption-line gas in our data (see Figure \ref{fig:ne_Corr}), so it is worth exploring the connection between the wind fluid and the gas we measure.  We begin by comparing the pressures implied by the densities we measure via \Siii\ to the pressure of the hot wind fluid predicted by CC85 at $r_*$. This pressure takes two forms, the thermal pressure of the wind fluid and its ram pressure. In convenient units, at $r_*$ the total (summed) pressure is given as:

\begin{equation}
P_{tot}/k = 1.79 \times 10^5 \text{SFR} \times r_*^{-2}\ [\text{K}\ \text{cm}^{-3}]
\end{equation}
where \text{SFR} is in units of \Msun/yr, $r_*$ is in units of kpc, and $k$ is the Boltzmann constant.
If we assume that the gas we measure with \Siii\ is in pressure equilibrium with the hot wind fluid, we have:

\begin{equation}
\label{eq:PressureEqui}
    P_\text{tot} = 2 \ne kT
\end{equation}
where \ne\ is the density of the outflows, the factor 2 of is due to the gas is highly ionized, and we assume T = 10,000 K, for the gas temperature. Thus, for pressure balance with the wind fluid, the gas densities traced by \Siii\ at $r_*$ are proportional to SFR/$r_*^2$ as:

\begin{equation}
\label{eq:PressureEqui2}
    \ne \simeq 9\times\text{SFR} \times r_*^{-2} [\text{cm}^{-3}]
\end{equation}

This allows us to compare the relationship predicted by the model to the data. We show this model as the blue line in the right panel of Figure \ref{fig:ne_Corr}. We see that the densities we measure are about an order-of-magnitude lower than the model. This suggests that we are measuring densities at radii significantly larger than $r_*$ where pressures are lower. 

In this region, the CC85 model shows that ram pressure is dominant over thermal pressure. Direct measurements of the radial density profiles for the optical emission-line gas show that this material is in pressure balance with the wind ram pressure \citep{Lehnert96}, so this is plausible for the ionized absorption-line gas as well. We can therefore compute the location (\Rram) at which the observed absorption-line gas is in pressure balance with the hot wind's ram pressure:

\begin{equation}
\label{eq:Pout}
    P_\text{ram} = \dot{p}_\text{SFR}/4 \pi r^2_\text{ram}
\end{equation}
where $P_\text{ram}$ is the wind ram pressure and $\dot{p}_\text{SFR}$ is the total momentum flux of the wind, which equals the input momentum from the starburst [reported in \cite{Xu22a} for our galaxies]. 


Combining Equations (\ref{eq:PressureEqui}) and (\ref{eq:Pout}), we can solve \Rram\ as:
\begin{equation}
\label{eq:RoutP}
    r_\text{ram} = \sqrt{\dot{p}_\text{SFR}/ (8 \pi \ne\ k T)}
\end{equation}

\begin{figure}
\center
	\includegraphics[page = 1, angle=0,trim={.0cm 0.1cm 0.1cm 0.3cm},clip=true,width=1.0\linewidth,keepaspectratio]{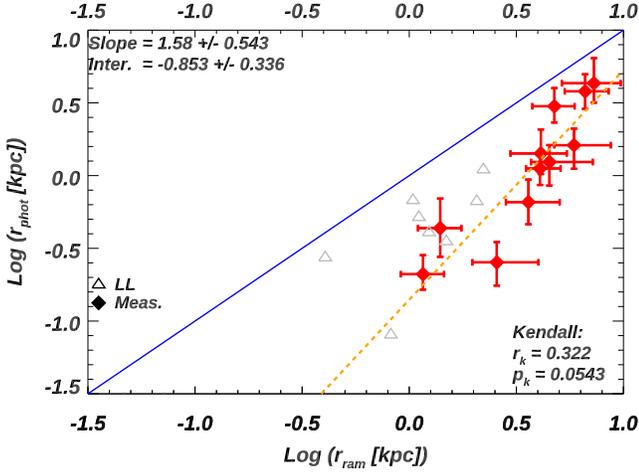}

\caption{\normalfont{Comparisons between outflow distances ($r$) derived from assuming photoionization (y-axis) and pressure equilibrium (x-axis). The labels and symbols are the same as Figure \ref{fig:ne_Corr}. We find the pressure-based sizes are  typically $\sim$ 2 to 5 times larger than the values derived from photoionization models, with the ratio decreasing from the small to large cases.} }
\label{fig:RComp}
\end{figure}

The derived results are shown in Table \ref{tab:mea}, and are in the range $\sim$ 1 to 8 kpc. In the right panel of Figure \ref{fig:RComp2}, we compare \Rram\ with $r_*$, which shows a strong correlation. In Figure \ref{fig:RComp}, we also compare \Rphot\ (Section \ref{sec:R_U}) with \Rram\ for each object. We find a linear relation with the pressure-based sizes being typically $\sim$ 5 times larger than the starburst radius.

\subsubsection{Outflow Sizes: Summary}

We have discussed three estimates of the radius of the outflow at the location at which the measured density occurs,based on different assumptions:

%
\begin{enumerate}
    \item The first is based on an outflow with a power-law radial density profile $n(r) = n_0 (r/r_{0})^{-\gamma}$. It predicts our measured densities occur at the characteristic radius $r_n =$ 1.73, 2.52, and 5.06$r_*$ for $\gamma$ = 2, 1.5, and 1.2 respectively (Section \ref{sec:Inter}). Here $r_*$ is the radius of the starburst.
 
    \item The second assumes the gas is photoionized by a fraction of the starburst ionizing flux that reaches the outflow. This estimates that the measured densities occur at a typical distance $\sim 1$ to $2 r_*$ (Section \ref{sec:R_U}). 
    
    
    \item Finally, we have assumed that the gas we measure is in pressure balance with the ram pressure of the hot wind fluid. This estimates that the measured densities occur at a typical distance $\sim 4 $ to $ 5 r_*$ (Section \ref{sec:R_P}).
    
\end{enumerate}

Given the systematic uncertainties in these estimates, we regard this level of agreement as satisfactory. In all cases, we are tracing the region of the outflow where densities are high enough to measure with our technique. As explained at the beginning of Section \ref{sec:R}, the maximum extent of the outflow could be considerably larger than \Rn.



\subsection{Other Important Outflow Parameters}
\label{sec:other}

Besides the density and structure of outflows, there are various essential parameters of outflows that have rarely been measured from observations. In this sub-section, we  constrain these parameters for outflows in our sample. We compare them with simulations of outflows and discuss the implications in Section \ref{sec:CompModel}.

We start with the volume filling factor (FF) of the observed outflow clouds, where we treat the absorbing material as an ensemble of clouds \citep[e.g.,][]{Fielding22}. This yields: 


\begin{equation}
\label{eq:FF}
    \text{FF} = \frac{N_\text{cl} \times 4/3 \pi R^{3}_\text{cl}}{A_\text{UV} r_n}
\end{equation}
where \Ncloud\ is the number of outflow clouds entrained in the hot wind at the outflow distance $r_n$, and $A_\text{UV}$ is the cross-sectional area of the starburst UV continuum. We also have the definition of outflow column density (\Nh) as:

\begin{equation}
\label{eq:Nh}
    \Nh\ = \frac{N_\text{cl} \times 4/3 \pi R^{3}_\text{cl} \times \nh}{A_\text{UV}}
\end{equation}

One can estimate FF from Equations (\ref{eq:FF}) and (\ref{eq:Nh}) as:

\begin{equation}
\label{eq:FF2}
    \text{FF} = \frac{\Nh }{\nh r_n}
\end{equation}
where, in this equation, all variables on the right side can be measured for at least part of the galaxies in our sample (see Sections \ref{sec:outflowSummary}, \ref{sec:ne_Corr}, and \ref{sec:R}). 

In \cite{Xu22a}, we also derived the area covering fraction of the outflow (CF) from the \Siii\ and \siiv\ absorption lines (Sections \ref{sec:outflowSummary}). We can rewrite CF as:

\begin{equation}
\label{eq:CF}
    \text{CF} = \beta_{sh} \times \frac{N_\text{cl} \times \pi R^{2}_\text{cl}}{A_\text{UV}}
\end{equation}
where \bsh\ is a coefficient between 0 and 1 to account for the shadowing effects. This is because the projected areas by different outflow clouds in the LOS can overlap each other so that their total covered area drops by the factor of \bsh. This factor depends on 1) the overall spatial distribution of outflow clouds; and 2) the second term of Equation (\ref{eq:CF}), i.e., the number and relative size of each cloud to A$_\text{UV}$. In Appendix \ref{sec:bsh}, we show how to estimate \bsh\ from Monte Carlo simulations and the measured CF in \cite{Xu22a}. For the 22 galaxies analyzed in this paper, we get \bsh\ in the range of $\sim$ 0.3 to 0.6. 


For simplicity of symbols, we define CF$_{sh}$ = CF/\bsh. From Equations (\ref{eq:FF}) and (\ref{eq:CF}), we can solve the size of the outflow clouds as:

\begin{equation}
\label{eq:R}
    R_\text{cl} = \frac{3}{4} \frac{\text{FF}}{\text{CF}_{sh}} r_{n}
\end{equation}

Using the above expression for $\text{FF}$ in Equation (\ref{eq:FF2}), this can be rewritten as:

\begin{equation}
\label{eq:CF2}
     R_\text{cl} = \frac{3}{4} \frac{\Nh}{\nh \text{CF}_{sh}}  
\end{equation}

This shows that $R_\text{cl}$ does not depend on $r_n$ and can be computed from directly measured quantities. Once we have $R_\text{cl}$ constrained, we can combine Equations (\ref{eq:CF}) and (\ref{eq:CF2}) to get \Ncloud\ as:

\begin{equation}
\label{eq:CF3}
N_\text{cl} = \text{CF}_{sh} \times \frac{A_\text{UV} }{\pi R^{2}_\text{cl}} 
            = \text{CF}_{sh} \times \frac{R^2_\text{UV} }{R^{2}_\text{cl}}
\end{equation}
where $R_\text{UV}$ is the UV size of a galaxy and we approximate it as \Rstar\ that we measured from the HST/COS acquisition images. Note that \Ncloud\ is also independent of $r_n$. We find the mean and median values of \Ncloud\ are 10$^{5.5}$ and 10$^{4.9}$, respectively.

Finally, we can estimate the average mass of the individual outflow clouds (\Mcloud) by:

\begin{equation}
\label{eq:Mcloud}
    M_\text{cl} = \frac{4}{3} \pi R^{3}_\text{cl} \nh \mu m_\text{p}
\end{equation}
where $\mu$ $\sim$ 1.4 is the average atomic mass per proton and $m_\text{p}$ is the proton mass.

We summarize the statistics for the derived FF, \Rcloud, \Mcloud\ values in Table \ref{tab:stat}, where their values for individual galaxy are shown in the last three columns in Table \ref{tab:mea}. For FF, which is the only derived parameter dependent on $r_n$, we have assumed $r_n$ = \Rram. If we assume $r_n =$ \Rphot, the derived \Rcloud\ and \Mcloud\ stay the same, while FF for each galaxy becomes larger with a mean and median value of 1.5\% and 13 pc, respectively. In Section \ref{sec:CompModel}, we compare these measurements with common outflow models, and discuss their implications.



\begin{figure*}
\center
    \includegraphics[page = 1, angle=0,trim={2.5cm 1.8cm 1.3cm 1.5cm},clip=true,width=0.5\linewidth,keepaspectratio]{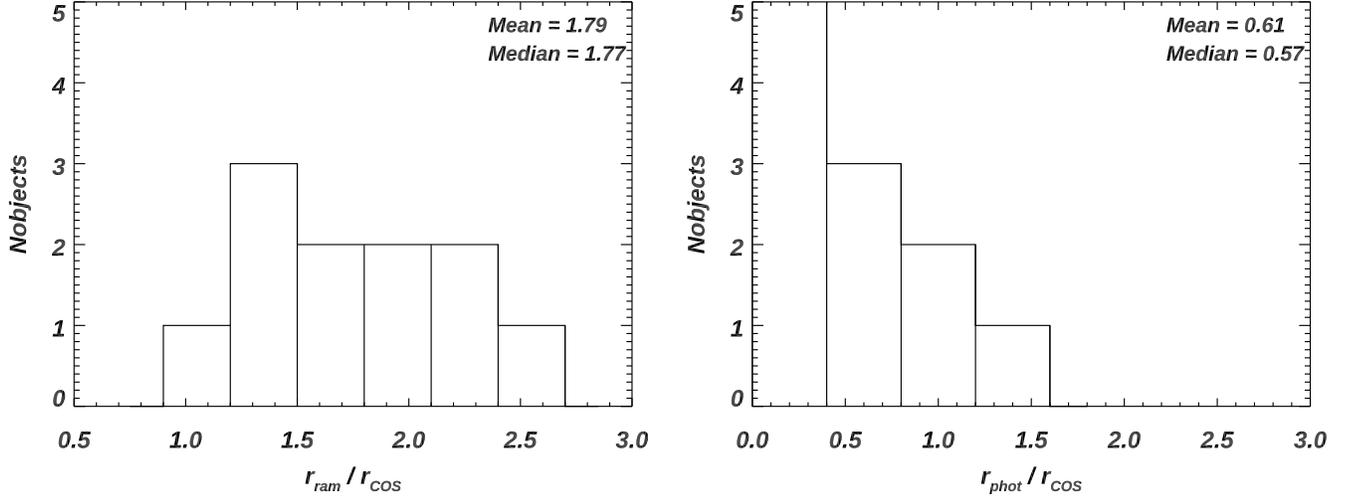}
    \includegraphics[page = 2, angle=0,trim={2.5cm 1.8cm 1.3cm 1.5cm},clip=true,width=0.5\linewidth,keepaspectratio]{./CLASSY_RvsRCOS.pdf}
\caption{\normalfont{Histograms showing the comparisons between the measured outflow radius (\Rram\ or \Rphot) and the projected physical size of the HST/COS aperture for each galaxy. The large ratio of \Rram/$r_\text{COS}$} suggests that the scattered or fluorescent emission lines should be weak in our galaxies, which is consistent with what has been found in the literature. See details in Section \ref{sec:CompDistance}.}
\label{fig:RComp3}
\end{figure*}

\section{Discussion}
\label{sec:Discussion}

\subsection{Comparisons with Other Outflow Density Measurements}

While we are presenting the first examples of density measurements for the warm ionized gas in outflows based on absorption lines, measurements of densities for the optical emission-line gas in outflows have been available in low-redshift starbursts for over thirty years \citep{Heckman90}. Here we summarize what has been learned from the optical emission-line gas and compare the results to our new data.

For low-redshift starbursts, \ne\ is commonly directly measured using the density-sensitive ratio of the [\sii] 6717 and 6731 emission lines. These data can be used to map out the radial variation in \ne, and show a steady radial decline from $\sim 500$ to 1000 cm$^{-3}$ in the starburst to $\sim$ 50 to 100 cm$^{-3}$ at distances several times larger than $r_*$ \citep[e.g.,][]{Heckman90,Lehnert96,Yoshida19,Perna20, Marasco22}. The [\sii] flux ratio reaches its low-density limit at $n_e \sim 10$ to $100$ cm$^{-3}$ \citep{Osterbrock06}, so direct measurements of \ne\ at larger radii (lower densities) are not possible. This would be consistent with the lower values of $n_e$ that we typically get in our sample, if we are probing larger radial scales in this case.

We have shown that with the assumption that the absorbing gas is in pressure balance with the ram pressure of the wind fluid, we do in fact derive large outflow radius. Is this a plausible assumption?  We believe it is, because both the emission- and absorption-lines trace the warm ionized gas phase. Since observations establish that the density (and pressure) profiles measured in the emission-line gas are consistent with the radial profile of the wind ram pressure \citep{Heckman90,Lehnert96}, this supports adopting the assumption of pressure balance to calculate $R_\text{ram}$ (Section \ref{sec:R_P}).


\subsection{Comparisons with Other Measurements of Outflow Structures}
\label{sec:CompDistance}
The outflow radii we derive assuming ram-pressure confinement are in the range $r_\text{ram} \sim$ 1 to 10 kpc. These size scales are consistent with those measured for the outflows traced by optical emission lines for starbursts with a range in SFR similar to our sample \citep[e.g.,][]{Armus95,Lehnert96,Martin98,Ho14, Yoshida19}. For the sample of dwarf SF galaxies in \cite{Marasco22}, which has weaker SFR than ours ($\sim$ 10 times smaller), they get $r$ $\sim$ 1 kpc. This is around the lower bound of our galaxies as expected from their lower SFR. 

Perhaps a more revealing comparison is to the size scales measured using resonantly-scattered emission arising from the same gas that produces the absorption lines seen directly along the LOS to the starburst \citep{Rubin11, Martin13}. This has been done recently using IFU instruments, including VLT/MUSE and Keck/KCWI, to map out \mgii\ emission lines surrounding starburst galaxies at intermediate redshifts \citep[e.g.,][]{Rupke19,Burchett21,Zabl21,Shaban22}. These data detect emission out to radii of $\sim$ 10 to 20 kpc, with half-light radii of 5 to 10 kpc. The latter is quite similar to values we derived for \Rram\ for our sample.

Given a typical aperture size of 1\arcsec\ -- 2\arcsec\ in current (non-IFU) spectrographs, these sizes imply that a significant fraction of the resonantly scattered line emission could lie outside the aperture. Thus, this missing light helps explain why the scattered (or fluorescently-reprocessed) emission-lines from the outflow are often quite weak \citep[e.g.,][]{Erb12, Steidel18, Wang20,Xu22a}, as can be seen in radiative transfer models of outflows \citep[e.g.,][and Huberty et al. in prep]{Prochaska11,Scarlata15,Carr21}. We explore this idea further with the current data. In Figure \ref{fig:RComp3}, we show histograms of the ratio $r_\text{ram}/r_\text{COS}$ and $r_\text{phot}/r_\text{COS}$, where $r_\text{COS}$ is the projected physical size of the COS aperture for a given galaxy.  We find that the ratios of $r_\text{ram}/r_\text{COS}$ are $>$ 1, while $r_\text{phot}/r_\text{COS}$ are most often $\lesssim$ 1. Thus, the larger sizes measured for $r_\text{ram}$ may be consistent with the relatively weak emission-lines seen in these galaxies \citep[e.g.,][]{Wang20,Xu22a}.

Another apt comparison is to maps of the outflows of neutral gas traced by the \nai\ D optical absorption-line \citep[1 -- 10 kpc, e.g.,][]{Martin06, Rupke13, Perna19, Avery21, Avery22}. Our data are complementary to these studies since they pertain to the ionized phase of the outflow and represent integrals over the line-of-sight directly into the starburst. Our data also provide information on key parameters like the densities, filling factors, radii and masses of the outflowing clouds.

Besides the outflows discussed above, SF galaxies can exhibit outflows features in many other wavelength bands and line diagnostics, where outflow distances are measured \citep[see reivews in][]{Heckman17a, Veilleux20}. These include very hot gas detected in X-ray \citep[at $\sim$ 1 -- 10 kpc, e.g.,][]{Strickland07, Li11, Zhang14}, and atomic and molecular outflows observed in infrared to radio bands \citep[e.g., from \protect{[\cii]} and CO, out to radii of a few kpc,][]{Walter02, Chisholm16c, Stuber21}. 

Again, we emphasize that these various outflow sizes are defined in different ways. In our case, we are defining the size to be the radius at which our measured densities occur. For the emission-line data, the sizes are typically just defined by the radius at which the emission becomes undetectably faint. Additionally, different diagnostics of outflows in different galaxies can reach intrinsically distinct scales, and the relationships between them are not entirely clear. Detailed comparisons are beyond the scope of this paper, but we plan to study the relationships between different diagnostics and phases of galactic outflows in future papers.






\subsection{Comparisons with Models and Simulations of Galactic Outflows}
\label{sec:CompModel}
Galactic winds are complex and difficult to model because one needs to simultaneously capture the large spatial scales for the whole galaxy and the fundamentally small-scale process happening between the galaxy's ISM/CGM and the wind \citep[see][and references therein]{Naab17}. Currently, a compelling model \citep[e.g.,][]{Fielding22} comprises 1) a hot, volume-filling wind component driven by thermalized ejecta of massive stars \citep{Chevalier85} \footnote{This hot gas is only detectable inside the starburst \citep{Heckman17a}, where its density is relatively high.} and 2) a cold to warm component in the form of embedded clouds, which are entrained by the hot wind. This component produces the observed outflows seen in UV absorption lines \citep[e.g.,][]{Xu22a}. The exchange of mass/momentum/energy between these two components is in the turbulent radiative mixing layer \citep[e.g.,][]{Gronke20, Tan21, Fielding22}. 
 
Two time-scales control the fate of the outflow clouds \citep{Gronke20}: 1) the clouds grow by cooling of the hot wind in a time scale of \tcool, which depends mainly on the pressure and metallicity; and 2) the clouds are destroyed by turbulent shredding in a time scale of \tmix. We have \tmix\ $\propto$ \Rcloud/\Vturb, where \Rcloud\ is the average radius of the outflow clouds and \Vturb\ is the turbulent velocity. For large outflow clouds, \tcool\ $<$ \tmix\ so that the clouds can grow. For smaller outflow clouds, the clouds are shredded before they can grow. Thus, parameters related to \Rcloud\ are important for galactic outflows but have rarely been constrained from observations. 

In Section \ref{sec:other}, we have shown that, based on our measurements of outflow density and distances, we can constrain these parameters, including FF, \Rcloud, and \Mcloud. Here we attempt to compare our measurements to common outflow models.

We can use the criterion derived by \citet{Gronke20} for the critical (minimum) size for a cloud to survive/grow when exposed to the ram pressure of the wind:
 
\begin{equation}
    \label{eq:Rcrit}
      R_{crit} \sim \frac{T_{cl,4}^{5/2} \mathcal{M}_{wind}}{P_3 \Lambda_{mix,-21.4}} \frac{\chi}{100} \alpha^{-1} pc
\end{equation}
 
Here, $T_{cl,4}$ is the cloud temperature in units of $10^{4}$ K, $\mathcal{M}_{wind}$ is the Mach number of the hot wind fluid, $P_{3}$ is the cloud pressure in units of $10^{3}$ K cm$^{-3}$, $\Lambda_{mix,-21.4}$ is the value of the cooling function in the turbulent mixing layer (in units of $10^{-21.4}$ cm$^{3}$ erg s$^{-1}$), $\chi$ is the ratio between the cloud and wind density, and $\alpha$ is a `fudge factor' of order unity. Under this model, if a cloud exposed to the hot wind has smaller sizes than \rcrit, it is destroyed/shredded before being accelerated.
  
We assume $T_{cl} = 10^4$ K and can then use our measured values of \ne\ to compute $P_3$. We further take $\alpha = 1$ and the fiducial value of $\Lambda_{mix}$. To measure $\chi$ we adopt the model above for clouds in pressure balance with the wind ram pressure. We use the \citet{Chevalier85} wind solution to obtain $\mathcal{M}_{wind}$ assuming \Rout\ = \Rram\ (see Section \ref{sec:R_P} and left panel of Figure \ref{fig:RComp2}). For balancing between the wind ram pressure and the cloud thermal pressure, we have 

\begin{equation}
P_{cl} = 2 n_{cl} k T_{cl} = \rho_w v_w^2
\end{equation}


Then, since $\rho_{cl} = n_{cl}  m_p$, we have:
 
\begin{equation}
     \chi = \frac{\rho_{cl}}{\rho_w} = \frac{v_w^2 m_p}{2 k T_{cl}}
\end{equation}

\begin{figure}
\center
	\includegraphics[page = 1, angle=0,trim={.0cm 0.1cm 0.1cm 0.3cm},clip=true,width=1.0\linewidth,keepaspectratio]{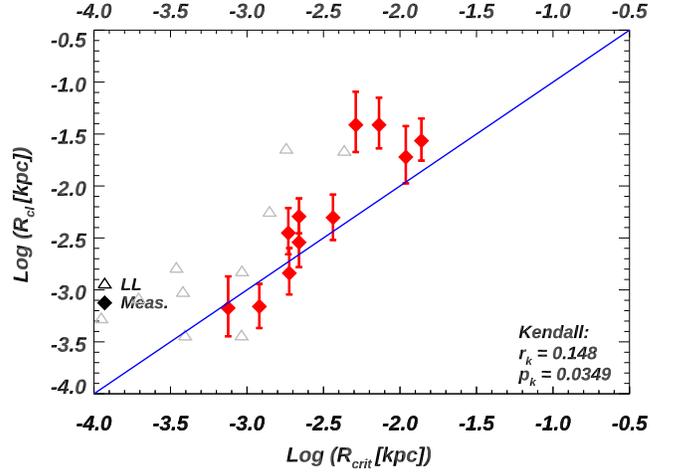}

\caption{\normalfont{Comparisons of the derived outflowing cloud radii (\Rcloud) with the critical (minimum) radius for a cloud to survive/grow when exposed to the ram pressure of the hot wind \citep{Gronke20}. The labels and symbols are the same as Figure \ref{fig:ne_Corr}. Galaxies that have \Rcloud\ as measurement or lower limits are shown as the red-filled or gray-open symbols, respectively. The blue line represents the 1:1 relationship. Within the uncertainties, 20/21 outflows have enough cloud sizes large enough to survive. See discussion in Section \ref{sec:CompModel}.} }
\label{fig:Rcl}
\end{figure}

Finally, we adopt the \citet{Chevalier85} model and assume a wind velocity of $v_w$ = 1800 km/s \citep{Strickland09}, leading to a value of $\chi \sim 2 \times 10^4$. We show the results in Figure \ref{fig:Rcl}, in which we compare our derived $R_\text{cl}$ with \rcrit\ for our sample. We find that the estimated values of $R_\text{cl}$ all lie close to \rcrit. Within the uncertainties, the growth criterion is satisfied in all 11 cases with \ne\ measurements. For the other 11 galaxies with \ne\ as upper limits, the derived values of \Rcloud\ and \rcrit\ are both lower limits (gray-open symbols). However, Equations (\ref{eq:CF2}) and (\ref{eq:Rcrit}) show that both sizes are inversely proportional to the density. This means that the ratio of \Rcloud/\rcrit\ is independent of density, and thus we can evaluate the growth criterion even in these cases. Within the uncertainties, 20 out of 21 cases\footnote{Among the total sample of 22 galaxies, one (J1612+0817) does not have \Nh\ reported in \cite{Xu22a} since its \siiv\ doublet troughs are in a detector gap of HST/COS. Thus, its \Rcloud/\rcrit\ ratio is unknown.} satisfy this criterion, i.e., $R_\text{cl}$ are large enough for them to survive under the impact of the hot wind. 

The fact that the cloud sizes are similar to \rcrit\ could be understood if the pre-existing population of clouds initially had a power-law distribution of sizes ($N_\text{cl} \propto R_\text{cl}^{-\gamma}$) that declines with increasing $R_\text{cl}$. Then only the clouds with $R_\text{cl} \gtrsim R_\text{crit}$ survive the interaction with the wind, while clouds with sizes $\gg$ \rcrit\ are rare (i.e., having a small total covering factor).

Additionally, in our assessment of cloud survival, Equations (\ref{eq:CF2}) and (\ref{eq:Rcrit}) imply that the ratio of \Rcloud/\rcrit\ depends only on the ratio of the column density to covering factor [since we adopted fixed values for all the terms in Equation (\ref{eq:Rcrit})]. Empirically, there is relatively small variation in CF$_{sh}$. In this case, the relatively small spread in the values of \Rcloud/\rcrit\ seen in Figure \ref{fig:Rcl} could imply that the total column densities of the absorbing clouds in the outflows are directly connected to the cloud-survival requirement. Future simulations of galactic winds may answer these implications.


\section{Conclusion and Future Work}
\label{sec:Conclusion}

We have reported here the first direct measurements of the density (\ne) in outflows from starburst galaxies traced by ultraviolet absorption lines. These measurements were made using COS on HST to measure the ratio of the column density of fine structure excited transitions of \Siii\ (i.e., \Siii*) to those of the \Siii\ resonance transitions. The sample of 22 galaxies was drawn from \cite{Berg22} and \cite{Heckman15}, and limited to cases with SNR $>$ 5, galaxy FUV radii $<$ 1.5\arcsec, and detected outflows. Our main results are as follows:

\begin{itemize}
    \item 
    We were able to measure \ne\ in 11 cases and set upper limits in the other 11 galaxies. The median density was 23 cm$^{-3}$. We found a strong correlation between \ne\ and the star-formation rate per unit area in the starburst.
    \item
    Since the value of \ne\ is derived along a line-of-sight, its meaning is only simple in the case of an expanding shell with constant density. In the case of a continuous outflow in which the density drops with radius, we showed that for radial density profiles with power-law indices of --2, --1.5, and --1.2, the measured densities would pertain to gas at respective radii of 1.7, 2.5, and 5.1 times the radius at which the outflow begins (taken to be the starburst radius).
    \item
    Using the measured values of \ne, we made two indirect estimates of the radius of outflows ($r_n$) at which this density applies. The first assumes that the gas is photo-ionized by radiation from the starburst. This required making estimates for the fraction of intrinsic ionizing radiation leaking out of the starburst and into the outflow. Typical radii from this method are 1 to 2 times the starburst radius. We then assumed that the absorbing gas clouds are in pressure equilibrium with the hot wind fluid. These radii are typically 4 to 5 times the starburst radius.
    \item
    We used the values of \ne\ and our measured values for the total hydrogen column density and the covering fraction of the outflow to estimate the radii and masses of the absorbing clouds. We found median values of $\sim$ 5 pc and 200 M$_{\odot}$ respectively. We also estimated the volume filling factor of the population of these clouds, with typical values of $10^{-3}$ to 10$^{-2}$. 
    \item
    We have compared the outflow clouds sizes to theoretical models in which clouds interact with a supersonic wind fluid. We find that in 20 out of 21 cases, the estimated cloud sizes exceed the critical cloud size, meaning that these clouds are predicted to survive and grow as they interact with a hot supersonic wind.
    
\end{itemize}

This is the first time that various essential absorption-line outflow parameters have been estimated from observations, including outflow density, volume filling-factor, and cloud sizes/masses. There are plenty of compelling future projects to do in both observations and simulations. For example, how do our derived \ne\ and $r_n$ values compare to direct measurements from spatially resolved observations? What are the differences and/or connections between the \ne\ and outflow sizes measured from emission and absorption line outflows? Given the detected warm outflows, can we provide constraints on the hot wind properties? How can the measured radii and masses of the absorbing clouds help constrain simulations of outflows? We have plans to tackle some of these questions in our future work.

\clearpage
\begin{acknowledgements}
X.X. and T.H. thank M. Gronke, D. Fielding, and G. Bryan for interesting discussions.

The CLASSY team is grateful for the support for this program, HST-GO-15840, that was provided by NASA through a grant from the Space Telescope Science Institute,  which is operated by the Associations of Universities for Research in Astronomy, Incorporated, under NASA contract NAS5- 26555. BLJ thanks support from the European Space Agency (ESA). CLM gratefully acknowledges support from NSF AST-1817125. The CLASSY collaboration extends special gratitude to the Lorentz Center for useful discussions during the "Characterizing Galaxies with Spectroscopy with a view for JWST" 2017 workshop that led to the formation of the CLASSY collaboration and survey.

Funding for SDSS-III has been provided by the Alfred P. Sloan Foundation, the Participating Institutions, the National Science Foundation, and the U.S. Department of Energy Office of Science. The SDSS-III web site is http://www.sdss3.org/.

SDSS-III is managed by the Astrophysical Research Consortium for the Participating Institutions of the SDSS-III Collaboration including the University of Arizona, the Brazilian Participation Group, Brookhaven National Laboratory, Carnegie Mellon University, University of Florida, the French Participation Group, the German Participation Group, Harvard University, the Instituto de Astrofisica de Canarias, the Michigan State/Notre Dame/JINA Participation Group, Johns Hopkins University, Lawrence Berkeley National Laboratory, Max Planck Institute for Astrophysics, Max Planck Institute for Extraterrestrial Physics, New Mexico State University, New York University, Ohio State University, Pennsylvania State University, University of Portsmouth, Princeton University, the Spanish Participation Group, University of Tokyo, University of Utah, Vanderbilt University, University of Virginia, University of Washington, and Yale University.

This research has made use of the HSLA database, developed and maintained at STScI, Baltimore, USA.

CHIANTI is a collaborative project involving George Mason University, the University of Michigan (USA), and the University of Cambridge (UK).


\end{acknowledgements}

\facilities{HST (COS)}
\software{
astropy \citep{Astropy22}
CalCOS (STScI),
jupyter \citep{Kluyver16},
MPFIT \citep{Markwardt09}}

\typeout{} 
\bibliography{main_MgII}{}
\bibliographystyle{aasjournal}

\appendix

\section{Estimations of \bsh}
\label{sec:bsh}
As discussed in Section \ref{sec:other}, different outflow clouds can ``shadow'' each other and produce smaller area covering fractions (CF) in the LOS \citep[e.g.,][]{Sun17,Xu20a}. Here we conduct a Monte Carlo (MC) experiment in two-dimension to estimate the shadowing parameter, \bsh, for Equation (\ref{eq:CF}).

\begin{figure}[H]
\center
	\includegraphics[page = 1, angle=0,trim={.0cm 0.1cm 0.1cm 0.3cm},clip=true,width=0.7\linewidth,keepaspectratio]{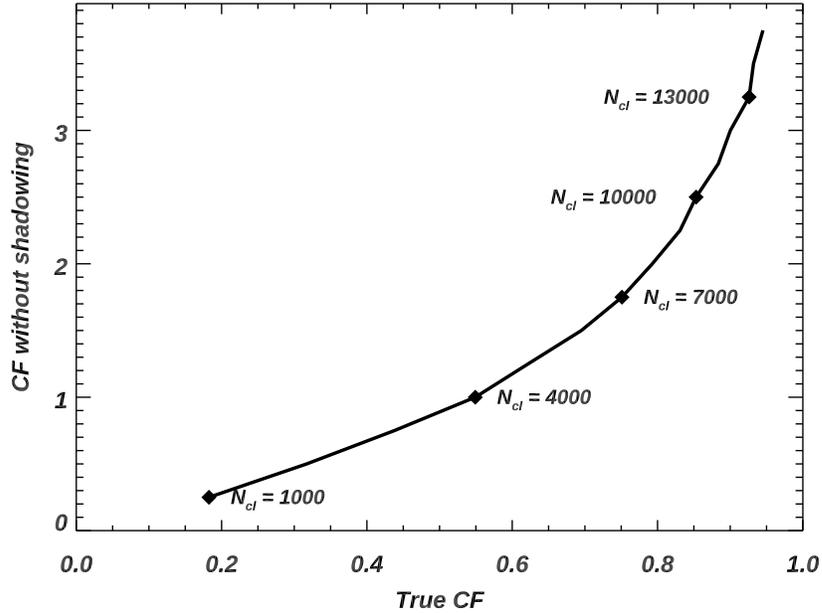}

\caption{\normalfont{Comparisons between CF assuming no shadowing versus the true CF from MC simulations. We also show the number of clouds (\Ncloud) for several positions on the line. See discussion in Appendix \ref{sec:bsh}.} }
\label{fig:MC}
\end{figure}

Given \Ncloud\ outflow clouds with radius \Rcloud, we randomly distributed them within the area of A$_\text{UV}$. We also assume the ratio of $A_\text{UV}/\pi R^{2}_\text{cl} =$ 4000. We have tested that the variations of this ratio have little effect on our final results below. We then vary \Ncloud\ from 1000 to 15,000 and calculate two quantities: 1) $N_\text{cl} \times \pi R^{2}_\text{cl}/A_\text{UV}$, which represents the CF value if outflow clouds do not shadow each other at all; and 2) the true CF by checking each spot in A$_\text{UV}$ to see if they are covered by any of the outflow clouds. We show these two quantities in the y- and x-axis in Figure \ref{fig:MC}, respectively. 

We find when \Ncloud\ grows, the true CF initially increases fast but then slows down. This is because when \Ncloud\ is small, we do not expect to have strong shadowing effects given the relatively large area of A$_\text{UV}$ compared to the projected size of each outflow cloud (i.e., $\pi R^{2}_\text{cl}$). But when \Ncloud\ is large ($\gtrsim$ 4000), the shadowing effects become more significant and the growth of the true CF is slower.

For galaxies in our sample, we have measured the true CF from ``down-the-barrel'' observations of UV absorption lines \citep{Xu22a}. Thus, we can estimate y = $N_\text{cl} \times \pi R^{2}_\text{cl}/A_\text{UV}$ for each galaxy based on the true CF and the curve in Figure \ref{fig:MC}. Then we can calculate \bsh\ from Equation (\ref{eq:CF}). For the 22 galaxies analyzed in this paper, we get \bsh\ in the range of $\sim$ 0.3 to 0.6.

\section{Tables}
Here we present the tables for the derived quantities for each galaxy.

\begin{turnpage}

\begin{table*}
	\centering
	\caption{Measured Parameters for Galaxies in the Combined Sample$^{(1)}$}
	\label{tab:mea}
	\begin{tabular}{llll llll llll lllll} 
		\hline
		\hline
		Object  &	log(SFR) & log(\Rstar) &  log(\Nh) & CF(\Siii)  & N(\Siii) & N(\Siii*) & log(\ne) & A$_\text{dust}$ & log(Q$_\text{eff}$) & log(\Rphot) & log(\Rram) & log(FF) & log(\RC) & log(\MC) &\bsh\\
		\hline
		        & \Msun/yr   & kpc           & cm$^{-2}$  & & 10$^{12}$cm$^{-2}$ & 10$^{12}$cm$^{-2}$  & cm$^{-3}$ & mags.  & s$^{-1}$ & kpc & kpc     &         & kpc      & \Msun &\\ 
		\hline
		(1)&(2)&(3)&(4)&(5)&(6)&(7)&(8)&(9)&(10)&(11)&(12)&(13)&(14)&(15)&(16)\\
		\hline
J0021+0052& 1.07& -0.05& 20.78& 0.30& 243.40& <62.55& <2.41& 1.81& 52.76& >-1.09& >-0.09& >-2.93& >-3.09& >1.20& 0.56& \\
J0150+1308$^\text{(H15)}$& 1.50& 0.10& 20.67& 0.85& 720.63& 18.89$^{+0.50}_{-0.50}$& 1.36$^{+0.04}_{-0.05}$& 2.81& 53.80& 0.09$^{+0.12}_{-0.16}$& 0.65$^{+0.20}_{-0.09}$& -2.73$^{+0.10}_{-0.21}$& -2.54$^{+0.24}_{-0.24}$& 1.81$^{+0.73}_{-0.72}$& 0.38& \\
J0808+3948& 1.26& -0.59& 19.99& 0.40& 741.47& <45.28& <1.73& 2.07& 54.10& >0.04& >0.35& >-3.49& >-3.45& >-0.54& 0.46& \\
J0823+2806& 1.48& -0.30& 20.56& 0.96& 187.19& 13.89$^{+4.50}_{-4.50}$& 1.82$^{+0.13}_{-0.18}$& 3.46& 53.09& -0.60$^{+0.14}_{-0.16}$& 0.41$^{+0.19}_{-0.11}$& -3.06$^{+0.22}_{-0.24}$& -3.18$^{+0.30}_{-0.27}$& 0.35$^{+0.92}_{-0.83}$& 0.34& \\
J0926+4427& 1.03& 0.12& 20.90& 0.30& 999.09& 7.93$^{+0.71}_{-0.71}$& 0.84$^{+0.13}_{-0.18}$& 1.21& 53.98& 0.48$^{+0.13}_{-0.11}$& 0.68$^{+0.09}_{-0.10}$& -2.01$^{+0.22}_{-0.18}$& -1.41$^{+0.26}_{-0.23}$& 4.67$^{+0.80}_{-0.70}$& 0.56& \\
J0938+5428& 1.05& 0.01& 20.86& 0.34& 1282.70& 14.40$^{+1.93}_{-1.93}$& 0.99$^{+0.15}_{-0.23}$& 1.00& 53.59& 0.15$^{+0.16}_{-0.14}$& 0.61$^{+0.12}_{-0.14}$& -2.14$^{+0.28}_{-0.20}$& -1.41$^{+0.32}_{-0.26}$& 4.81$^{+0.97}_{-0.82}$& 0.61& \\
J1016+3754& -1.17& -0.61& 20.74& 0.25& 44.12& <18.21& <2.65& 0.68& 51.99& >-1.54& >-1.33& >-1.97& >-3.29& >0.85& 0.59& \\
J1024+0524& 0.21& -0.27& 21.00& 0.62& 384.52& <59.64& <2.16& 0.90& 53.09& >-0.56& >-0.39& >-2.16& >-2.80& >1.81& 0.48& \\
J1025+3622& 1.04& 0.20& 20.84& 0.69& 612.97& 2.57$^{+0.67}_{-0.67}$& 0.56$^{+0.11}_{-0.14}$& 1.65& 53.83& 0.58$^{+0.12}_{-0.12}$& 0.82$^{+0.11}_{-0.10}$& -1.93$^{+0.17}_{-0.16}$& -1.56$^{+0.21}_{-0.19}$& 3.95$^{+0.65}_{-0.59}$& 0.38& \\
J1144+4012& 1.51& 0.26& 20.71& 0.92& 1482.90& 23.40$^{+1.06}_{-1.06}$& 1.14$^{+0.04}_{-0.05}$& 2.92& 53.68& 0.21$^{+0.11}_{-0.16}$& 0.77$^{+0.17}_{-0.09}$& -2.59$^{+0.11}_{-0.18}$& -2.30$^{+0.22}_{-0.21}$& 2.31$^{+0.67}_{-0.65}$& 0.37& \\
J1148+2546& 0.53& 0.37& 21.06& 0.87& 1615.60& <21.59& <1.06& 2.98& 52.76& >-0.18& >0.31& >-1.71& >-1.67& >4.10& 0.47& \\
J1150+1501& -1.33& -0.87& 20.71& 0.71& 380.18& <53.65& <2.12& 2.07& 51.23& >-1.54& >-1.14& >-1.65& >-3.04& >1.07& 0.49& \\
J1200+1343& 0.75& -0.35& 20.85& 0.84& 789.16& <47.91& <1.73& 2.58& 53.17& >-0.39& >0.09& >-2.37& >-2.83& >1.30& 0.33& \\
J1253--0312& 0.56& -0.11& 21.00& 0.91& 404.83& 12.42$^{+2.38}_{-2.38}$& 1.43$^{+0.12}_{-0.16}$& 2.51& 52.96& -0.36$^{+0.20}_{-0.20}$& 0.14$^{+0.10}_{-0.10}$& -1.96$^{+0.20}_{-0.16}$& -2.44$^{+0.24}_{-0.21}$& 2.17$^{+0.73}_{-0.64}$& 0.29& \\
J1359+5726& 0.42& 0.17& 21.05& 0.81& 979.65& <39.70& <1.55& 1.35& 53.31& >-0.17& >0.02& >-1.91& >-2.26& >2.85& 0.43& \\
J1416+1223& 1.57& -0.26& 20.27& 0.54& 175.12& 8.24$^{+0.71}_{-0.71}$& 1.62$^{+0.07}_{-0.09}$& 2.20& 53.63& -0.18$^{+0.16}_{-0.15}$& 0.56$^{+0.15}_{-0.11}$& -3.29$^{+0.14}_{-0.17}$& -3.15$^{+0.22}_{-0.21}$& 0.22$^{+0.66}_{-0.63}$& 0.40& \\
J1428+1653& 1.22& 0.32& 20.70& 0.62& 436.32& 2.32$^{+0.77}_{-0.77}$& 0.66$^{+0.14}_{-0.20}$& 1.04& 54.26& 0.63$^{+0.17}_{-0.13}$& 0.86$^{+0.13}_{-0.15}$& -2.21$^{+0.26}_{-0.19}$& -1.72$^{+0.30}_{-0.25}$& 3.58$^{+0.90}_{-0.79}$& 0.43& \\
J1429+0643& 1.42& -0.06& 20.81& 0.67& 675.03& 17.69$^{+0.78}_{-0.78}$& 1.36$^{+0.06}_{-0.07}$& 1.58& 53.80& 0.05$^{+0.09}_{-0.11}$& 0.61$^{+0.10}_{-0.06}$& -2.55$^{+0.11}_{-0.13}$& -2.29$^{+0.17}_{-0.16}$& 2.55$^{+0.53}_{-0.49}$& 0.44& \\
J1448--0110& 0.39& -0.60& 20.50& 0.83& 106.60& 3.22$^{+0.92}_{-0.92}$& 1.42$^{+0.12}_{-0.17}$& 2.76& 52.45& -0.68$^{+0.13}_{-0.11}$& 0.06$^{+0.10}_{-0.10}$& -2.38$^{+0.20}_{-0.16}$& -2.84$^{+0.24}_{-0.20}$& 0.96$^{+0.73}_{-0.64}$& 0.34& \\
J1545+0858& 0.37& -0.31& 21.38& 0.61& 1435.50& <45.45& <1.44& 1.55& 53.12& >-0.29& >0.05& >-1.50& >-1.66& >4.53& 0.51& \\
J1612+0817& 1.58& 0.01& \dots& 0.80& 2665.30& <230.03& <1.89& 2.75& 53.83& \dots& >0.42& \dots& \dots& \dots& \dots& \\
J2103--0728$^\text{(H15)}$& 1.29& -0.32& 20.47& 0.84& 1526.10& <209.23& <2.10& 4.18& 53.43& >-0.45& >0.17& >-3.20& >-3.44& >-0.16& 0.40& \\

		\hline
		\hline
	\multicolumn{16}{l}{%
  	\begin{minipage}{23cm}%
		Notes: (1)\ \ Measured parameters for 22 galaxies in our combined sample that have high SNR spectra of \Siii\ and \Siii*\ (see Section \ref{sec:mea}). Galaxies from \cite{Heckman15} is marked as (H15). Descriptions for each column: (2) -- (6) are adopted from \cite{Xu22a} (see Section \ref{sec:outflowSummary} for a summary); (2) The log of star-formation rate (SFR) of the galaxy; (3) The log of starburst radius of the galaxy, where we take \Rstar\ = 2 $\times r_\text{50}$ from \cite{Xu22a}; (4) The log of total hydrogen column density of the outflow; (5) The mean covering fraction derived from \Siii\ outflow absorption lines; (6) The column density of \Siii\ in the outflows; (7) The column density of excited states of \Siii\ (i.e., \Siii*) measured in the outflow absorption troughs for each galaxy (Section \ref{sec:mea}); (8) The log of electron number density of the outflows (Section \ref{sec:mechan}); (9) Dust extinction derived from SED fittings (Section \ref{sec:R_U}); (10): The effective ionizing photon rate per second for the outflows (Section \ref{sec:R_U}); (11) The radius of the observed outflows derived from photoionization models (Section \ref{sec:R_U}); (12) The radius of the observed outflows derived from assuming pressure equilibrium (Section \ref{sec:R_P}); (13) The volume filling factor for outflows (Section \ref{sec:other}); (14) and (15) The average radius and mass of the outflowing clouds (Section \ref{sec:other}) adopting \Rram\ (Section \ref{sec:other}). (16) The coeffcient between 0 --1 to account for the shadowing effects of outflow clouds (see Section \ref{sec:other} and Appendix \ref{sec:bsh})\\
  	\end{minipage}%
	}
	\end{tabular}
	\\ [0mm]
	
\end{table*}

\end{turnpage}

\clearpage

\end{document}